\documentclass[conference]{IEEEtran}

\usepackage{array}
\usepackage{fixltx2e}
\usepackage{stfloats}
\usepackage{todo}

\usepackage[backend=biber, style=ieee]{biblatex}
\usepackage{url}
\usepackage{booktabs}
\usepackage{svg}
\usepackage{graphicx}
\usepackage{float}
\usepackage{threeparttable}

\usepackage{lscape}
\usepackage{tabu}
\usepackage{longtable}
\usepackage{csquotes}
\begin{document}
\title{SOK: Building a Launchpad for Impactful Satellite Cyber-Security Research}

\author{\IEEEauthorblockN{James Pavur}
\IEEEauthorblockA{Oxford University \\
james.pavur@cs.ox.ac.uk}
\and
\IEEEauthorblockN{Ivan Martinovic}
\IEEEauthorblockA{Oxford University\\
ivan.martinovic@cs.ox.ac.uk}}
%\and
%\IEEEauthorblockN{James Kirk\\ and Montgomery Scott}
%\IEEEauthorblockA{Starfleet Academy\\
%someemail@somedomain.com}}


%\author{\IEEEauthorblockN{James Pavur\IEEEauthorrefmark{1},
%		Daniel Moser\IEEEauthorrefmark{2},
%		Martin Strohmeier\IEEEauthorrefmark{2}, 
%		Vincent Lenders\IEEEauthorrefmark{2} and
%		Ivan Martinovic\IEEEauthorrefmark{1}}
%	\IEEEauthorblockA{\IEEEauthorrefmark{1}Oxford University\\ Email: 
%\textit{first.last}@cs.ox.ac.uk}
%	\IEEEauthorblockA{\IEEEauthorrefmark{2}armasuisse\\
%		Email: \textit{first.last}@armasuisse.ch}}

\IEEEoverridecommandlockouts
\makeatletter\def\@IEEEpubidpullup{6.5\baselineskip}\makeatother
\IEEEpubid{\parbox{\columnwidth}{
	}
	\hspace{\columnsep}\makebox[\columnwidth]{}}

\maketitle

\begin{abstract}
	As the space industry approaches a period of rapid change, securing both emerging and legacy satellite missions will become vital. However, space technology has been largely overlooked by the systems security community. This systematization of knowledge paper seeks to understand why this is the case and to offer a starting point for technical security researchers seeking impactful contributions beyond the Earth's mesosphere.
	
	The paper begins with a cross-disciplinary synthesis of relevant threat models from a diverse array of fields, ranging from legal and policy studies to aerospace engineering. This is presented as a ``threat matrix toolbox'' which security researchers may leverage to motivate technical research into given attack vectors and defenses. We subsequently apply this model to an original chronology of more than 100 significant satellite hacking incidents spanning the previous 60 years. Together, these are used to assess the state-of-the-art in satellite security across four sub-domains: satellite radio-link security, space hardware security, ground station security, and operational/mission security. In each area, we note significant findings and unresolved questions lingering in other disciplines which the systems security community is aptly poised to tackle. By consolidating this research, we present the case that satellite systems security researchers can build on strong, but disparate, academic foundations and rise to meet an urgent need for future space missions.
\end{abstract}

\section{Introduction}
From the launch of Sputnik in October of 1957, space technology has played a critical role in the emergence of the information age. Today, satellites are far more than simple scientific demonstrations, instead underpinning essential services that define our lives. As the satellite industry undergoes a market renaissance driven by miniaturization and declining launch costs, understanding and defending these systems against cyber-attacks can only increase in importance.

Rather than presenting a direct survey of satellite trends and emerging security technologies, something well-provided in prior work, this paper is motivated by an enduring problem for space systems security research~\cite{manulisCyberSecurityNew2020}. In the status quo, satellite cyber-security is a disparate and ill-defined topic, with critical contributions scattered across diverse disciplines ranging from history and security studies to aerospace engineering and astrophysics. Academics in each domain have made valuable discoveries, but contributions in one field are easily overlooked by researchers with a narrow focus on their own.

This paper offers a cross-disciplinary synthesis of progress to date on space systems security. The paper begins by presenting a unified matrix of existing threat models - linking attackers, vulnerabilities and motivations drawn from dozens of prior studies. Underpinning this effort is an exhaustive historical timeline of satellite hacking incidents, where our own archival research is added to prior contributions from Fritz and Manulis et al.~\cite{fritzSatelliteHackingGuide2013,manulisCyberSecurityNew2020}. The combination of this historical analysis and threat modeling framework offers a useful aid to those seeking credible and empirical threat models as motivation for systems security research on satellites.

Beyond this, we further analyze these historical incidents through the lens of four main problem domains: RF-link security, space platform security, ground systems security, and mission operations security. In each, we not only highlight relevant work and trends, but draw out key unsolved questions from other fields which the systems security community is aptly situated to tackle. The ultimate motivation for the paper is to provide a launchpad for technical security researchers seeking unique and enduring challenges in space.

\section{Threat Modeling in Context}
In order to identify needs which may be fulfilled by the systems security community, a robust understanding of the means and motivations of attackers is necessary. This section contextualizes previous work in light of emerging industry trends to arrive at a high-level model of threats and vulnerabilities impacting space systems.

\subsection{The Rise of Satellites} 
Today, more than 2,000 operational satellites orbit Earth, supporting a market worth more than \$150 billion annually~\cite{unionofconcernedscientistsUCSSatelliteDatabase2018,dscSpaceFinalFrontier2016}. They underpin a wide range of vital services, including: more than 10~TB/s of global internet capacity, media broadcasts over 100 million customers, daily terabytes of earth observation data, and precise global positioning services~\cite{dscSpaceFinalFrontier2016}. Their importance will only increase. By 2035, satellite broadband is anticipated to exceed 100~TB/s globally and the direct industry value will exceed half a trillion dollars annually~\cite{dscSpaceFinalFrontier2016}.

Around 40\% of operational satellites are used for business communications and 30\% support a mix of civilian and military government operations, the remainder are dedicated to mixed-use remote-sensing, meteorological, and navigational missions~\cite{bardinSatelliteCyberAttack2013}. However, this balance will likely shift in response to demand for ubiquitous broadband service and remote sensing capacity. The emerging sector rising to meet this demand is widely referred to as ``New Space''~\cite{weinzierlSpaceFinalEconomic2018}. Among the most prominent New Space missions are mega-constellations proposed by organizations like Blue Origin, SpaceX, and OneWeb. If successful, these projects will increase the number of Low Earth Orbit (LEO) satellites by an order of magnitude.

The most important driver of these changes is diminishing launch costs. Modern launch vehicles have reduced to cost-per-kilogram to LEO to under \$2,000~\cite{tartarWhichRocketsAre2018}. This is radically more affordable than NASA's famous shuttle missions (at around  \$54,500), and almost 90\% cheaper than the average cost of all missions from 1970-2000 (around \$18,500)~\cite{jonesRecentLargeReduction2018}. For the first time, the deployment of satellite payloads is within the means of a vast array of new industry entrants.

Concurrent improvements in computing capabilities - particularly with respect to miniaturization - have compounded these effects. As computer hardware grows smaller and less power-demanding, increasingly complex and light satellites become feasible. This has resulted in the emergence of ``small satellites'' - a wide range of sub-500~kg devices, with many weighing less than 1~kg.

The emergence of commercial off-the-shelf (COTS) satellite component has further driven growth in the small satellite market. The availability of ready-made satellite flight hardware decreases procurement costs, allowing New Space entrants to accept larger technical and commercial risks. Indeed, it is now possible to purchase a fully assembled 1~kg ``Cube Satellite'' for as little as \$16,000~\cite{interorbitalInterorbitalStorefront}.

\subsection{Emerging Threat Landscape}
\label{sec:emerging-threats}
As the demand for and usage of space assets grows, the threat environment they face has shifted. Historically, satellites have benefited from a sort of ``security through obscurity'' whereby system complexity and equipment costs dissuade all but the most sophisticated cyber-adversaries. The combined effects of COTS components and constellations with thousands of identical satellites mean that diversity and complexity of implementation are unlikely to provide enduring security.

In general, the threat to satellites is well understood and intuitive. In a military context, space systems are essential for Command, Control, Communications, Computer, Intelligence, Surveillance, and Reconnaissance (C4ISR) capabilities~\cite{grantSpaceDependenceCritical2005, lungermanWhatHappensIf2014}. As a result, adversaries seeking to ``level the playing field'' against great powers have strong incentives to target satellites~\cite{pavurCyberASATImpactCyber2019}. Civil society also depends heavily on space services, whether those take the form  of positioning data essential to modern transport and logistics or meteorological services which protect millions from natural disasters. Those seeking to cause societal disruption may perceive satellites as an attractive ``single point of failure'' in many critical infrastructures~\cite{falcoVacuumSpaceCyber2018}.

\begin{table}
	\centering
	\caption{Summary of Satellite Threat Actors}
	\label{tab:attacker_types}
	\resizebox{\linewidth}{!}{%
		\begin{tabular}{lll} 
			\toprule
			\begin{tabular}[c]{@{}l@{}}\textbf{Attacker}\\\textbf{Type} \end{tabular} & \begin{tabular}[c]{@{}l@{}}\textbf{Example }\\\textbf{Motivations} \end{tabular}                                                                                                    & \begin{tabular}[c]{@{}l@{}}\textbf{Technical }\\\textbf{Capabilities} \end{tabular}  \\ 
			\midrule
			\begin{tabular}[c]{@{}l@{}}National\\Military\end{tabular}                & \begin{tabular}[c]{@{}l@{}}\begin{tabular}{@{\labelitemi\hspace{\dimexpr\labelsep+0.5\tabcolsep}}l}Space Control\\Anti-Satellite Weapons \end{tabular}\end{tabular}                 & High                                                                                 \\ 
			\hline
			\begin{tabular}[c]{@{}l@{}}State\\Intelligence\end{tabular}               & \begin{tabular}[c]{@{}l@{}}\begin{tabular}{@{\labelitemi\hspace{\dimexpr\labelsep+0.5\tabcolsep}}l}Counter-Intelligence\\Technology Theft\\Eavesdropping \end{tabular}\end{tabular} & High                                                                                 \\ 
			\hline
			\begin{tabular}[c]{@{}l@{}}Industry\\Insiders \end{tabular}               & \begin{tabular}[c]{@{}l@{}}\begin{tabular}{@{\labelitemi\hspace{\dimexpr\labelsep+0.5\tabcolsep}}l}Sabotage\\Technology Theft\end{tabular}\end{tabular}                             & High                                                                                 \\ 
			\hline
			\begin{tabular}[c]{@{}l@{}}Parts\\Suppliers \end{tabular}                 & \begin{tabular}[c]{@{}l@{}}\begin{tabular}{@{\labelitemi\hspace{\dimexpr\labelsep+0.5\tabcolsep}}l}Sabotage\\Espionage \end{tabular}\end{tabular}                                   & High                                                                                 \\ 
			\hline
			\begin{tabular}[c]{@{}l@{}}Organized \\Crime \end{tabular}                & \begin{tabular}[c]{@{}l@{}}\begin{tabular}{@{\labelitemi\hspace{\dimexpr\labelsep+0.5\tabcolsep}}l}Eavesdropping\\Ransom\\Technology Theft\end{tabular}\end{tabular}                & Moderate                                                                             \\ 
			\hline
			\begin{tabular}[c]{@{}l@{}}Commercial\\Competitors \end{tabular}          & \begin{tabular}[c]{@{}l@{}}\begin{tabular}{@{\labelitemi\hspace{\dimexpr\labelsep+0.5\tabcolsep}}l}Sabotage \\Technology Theft\end{tabular}\end{tabular}                            & Moderate                                                                             \\ 
			\hline
			Terrorists                                                                & \begin{tabular}[c]{@{}l@{}}\begin{tabular}{@{\labelitemi\hspace{\dimexpr\labelsep+0.5\tabcolsep}}l}Societal Harm\\Notoriety\\Message Broadcasting \end{tabular}\end{tabular}        & Low                                                                                  \\ 
			\hline
			Individuals                                                               & \begin{tabular}[c]{@{}l@{}}\begin{tabular}{@{\labelitemi\hspace{\dimexpr\labelsep+0.5\tabcolsep}}l}Notoriety\\Personal Challenge \end{tabular}\end{tabular}                         & Low                                                                                  \\ 
			\hline
			\begin{tabular}[c]{@{}l@{}}Political\\Activists \end{tabular}             & \begin{tabular}{@{\labelitemi\hspace{\dimexpr\labelsep+0.5\tabcolsep}}l}Message Broadcasting \end{tabular}                                                                          & Low                                                                                  \\
			\bottomrule
		\end{tabular}
	}
\end{table}

With regards to potential attackers, a 2016 report by Chatham House, a prominent UK policy think-tank, taxonomizes threat actors into four broad categories: states seeking military advantage, organized criminal efforts for financial gain, terrorist groups seeking recognition, and individual hackers proving their skills~\cite{dscSpaceFinalFrontier2016}. This can be supplemented with the list of threat actors published by the Consultative Committee for Space Data Systems (CCSDS)~\cite{ccsdsSecurityThreatsSpace2015}. CCSDS represents a consortium of national space agencies from eleven member states and thirty-two observer nations and is one of the most influential technical bodies for the development of space protocol and systems standards. Beyond overlaps with Chatham House's model, CCSDS adds: foreign intelligence services, political activists, commercial competitors, agency insiders and business partners~\cite{ccsdsSecurityThreatsSpace2015}. Independent authors within the military strategy and civil space science domains have further suggested supply-chain threats from equipment manufacturers~\cite{laneHighAssuranceCyberSpace2017, delmonteCybersecurityPolicySustainable2013}. Table~\ref{tab:attacker_types} offers a composite summary of threat actors from these and other reports as a starting point for the development of threat models~\cite{harrisonSpaceThreatAssessment2018,fidlerCybersecurityNewEra2018}.

It is worth noting that our research has been restricted to English-language resources, which tend to show a western bias in threat. For example, the Center for Strategic and International Studies (CSIS), a Washington DC political and security think-tank, isolates four main state belligerents in orbit: Russia, China, Iran and North Korea~\cite{harrisonSpaceThreatAssessment2018}. CSIS contends that Russian cyber-capabilities against satellites are particularly sophisticated and have been demonstrated in historical attacks on critical infrastructure and space systems. With respect to China, CSIS highlights the fact that the People's Liberation Army Strategic Support Force (SSF) has organizational responsibility over both China's counterspace weapons and offensive cyber operations - creating natural cross-over opportunities. Other sources note that Chinese military reports have explicitly advocated for the use of digital counterspace against US space assets~\cite{fidlerCybersecurityNewEra2018}. Less information is provided to motivate the North Korean and Iranian threat, but CSIS notes sustained Iranian interest in cyber-attacks against the related ballistic missile defense (BMD) domain and North Korean cyber-attacks against terrestrial critical infrastructure. Very few English-language sources offer deep threat assessments of US and EU offensive capabilities in space, but it is perhaps not unreasonable to assume similar interests and means.

\subsection{Vulnerability Classes}

In addition to understanding \textit{who} might be interested in harming satellites, it is important to consider \textit{how} they might go about doing so. A high-level starting point can be found in the security studies and international relations fields, where scenario modeling is a common component of strategic analysis. Chatham House groups threats to satellites into two broad categories: attacks which target satellites themselves (e.g. via control system exploitation) and attacks which target satellite ground stations (e.g. via traditional network intrusion)~\cite{dscSpaceFinalFrontier2016}. The European Space Agency (ESA) brings civilian governmental perspectives, outlining additional threats to scientific missions including signal intercept and jamming, denial of service attacks, and supply chain malware~\cite{delmonteCybersecurityPolicySustainable2013}. Further technical specifics can be gleaned from CCSDS which adds replay attacks, access-control failures, social engineering, data corruption, and meta-data analysis on encrypted traffic~\cite{ccsdsSecurityThreatsSpace2015}. In their research on the intersection between space and military law, Rendleman and Ryals raise the novel additional threat of satellite hijackers who steal orbiting satellites to bolster their own space capabilities~\cite{rendlemanCyberOperationsDefend2013}. Finally, multiple researchers across the communications and systems security domain have considered the threat of signal piracy and spoofing attacks~\cite{laneHighAssuranceCyberSpace2017,Tippenhauer:2011:RSG:2046707.2046719,fritzSatelliteHackingGuide2013}.

\begin{figure}[h]
	\includesvg[width=\linewidth, inkscapelatex=false]{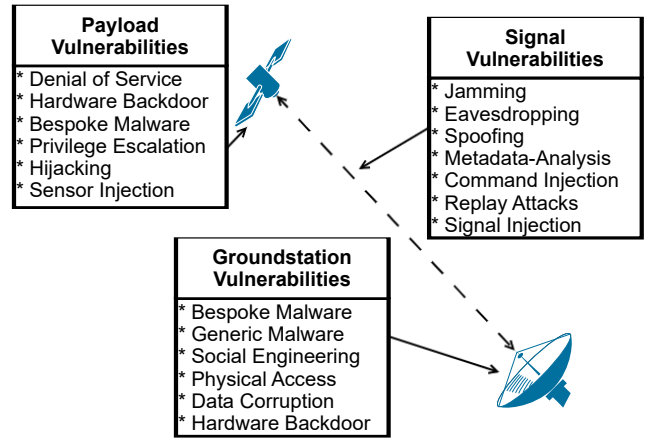}
	\caption{A Subsystem Approach To Satellite Vulnerability Classification}
	\label{fig:attack-taxonomy}
\end{figure}

We can bring structure to this diverse array of perspectives by expanding Chatham House's systemic taxonomy slightly. Specific, we propose three broad categories of attack surfaces: those relating to satellite signals, those relating to the space-platform and those which target satellite ground systems~(Figure~\ref{fig:attack-taxonomy}). Some alternative models will further divide ground systems into ``Customer'' and ``Mission'' segments, but we find that most threat models impact both use-cases~\cite{manulisCyberSecurityNew2020}.

While our classification system still results in some overlap, such as the case where RF-signals are used to send malicious flight commands to space platform, it has two key benefits. First, it aligns closely with common organization paradigms around space missions. Satellite missions are multi-stakeholder processes, where distinct organizations are often responsible for the on-orbit operations, communications, and ground segments. By mapping vulnerabilities to these domains, we can better clarify which organization has responsibility for defending against which threats. A second benefit of this approach is that the technical skills required for systems security research in each domain are intuitively distinct: on-orbit defenses draw from the embedded and control systems topics; signals defense requires networking and radio expertise; and ground systems leverage traditional operational technology (OT) and information technology (IT) perspectives.

We further can combine this subsystem taxonomy with the threat actors outlined in Section~\ref{sec:emerging-threats}. This allows us to develop a matrix synthesizing prior work into a mapping of threat actors and capabilities to vulnerabilities and impacted subsystems as shown in Table~\ref{tab:threats_to_attacks}.

\begin{table*}[t]
	\centering
	\caption{Demonstrative Matrix of Threat Actors, Capabilities, Objectives, and Vulnerabilities}
	\label{tab:threats_to_attacks}
	\resizebox{\linewidth}{!}{%
		\begin{threeparttable}
			\begin{tabular}{lllccccccccc} 
				\toprule
				\begin{tabular}[c]{@{}l@{}}\textbf{Vulnerability}\\\textbf{Type}\end{tabular} & \begin{tabular}[c]{@{}l@{}}\textbf{Example}\\\textbf{Attack}\end{tabular} & \begin{tabular}[c]{@{}l@{}}\textbf{Relevant}\\\textbf{Subsystems}\end{tabular} & \multicolumn{1}{l}{\textbf{Military} } & \multicolumn{1}{l}{\begin{tabular}[c]{@{}l@{}}\textbf{Intelligence}\\\textbf{Agency}\end{tabular}} & \multicolumn{1}{l}{\begin{tabular}[c]{@{}l@{}}\textbf{Corporate}\\\textbf{Insider} \end{tabular}} & \multicolumn{1}{l}{\begin{tabular}[c]{@{}l@{}}\textbf{Hardware}\\\textbf{Supplier}\end{tabular}} & \multicolumn{1}{l}{\begin{tabular}[c]{@{}l@{}}\textbf{Organized}\\\textbf{Crime}\end{tabular}} & \multicolumn{1}{l}{\begin{tabular}[c]{@{}l@{}}\textbf{Corporate}\\\textbf{Competitor}\end{tabular}} & \multicolumn{1}{l}{\begin{tabular}[c]{@{}l@{}}\textbf{Terrorist}\\\textbf{Group}\end{tabular}} & \multicolumn{1}{l}{\begin{tabular}[c]{@{}l@{}}\textbf{Individual} \\\textbf{Hacker}\end{tabular}} & \multicolumn{1}{l}{\begin{tabular}[c]{@{}l@{}}\textbf{Activist}~\\\textbf{Group}\end{tabular}}  \\ 
				\cmidrule{1-12}
				\begin{tabular}[c]{@{}l@{}}Denial of \\Service\end{tabular}                   & \begin{tabular}[c]{@{}l@{}}Forced "Safe\\~Mode"\end{tabular}              & Payload                                                                        & \checkmark                                    & c                                                                                                  & \checkmark                                                                                               & c                                                                                                & \checkmark                                                                                            & i                                                                                                   & i                                                                                              & i                                                                                                 & x                                                                                               \\ 
				\hline
				\begin{tabular}[c]{@{}l@{}}Hardware\\Backdoor\end{tabular}                    & \begin{tabular}[c]{@{}l@{}}Malicious Bus \\Messages\end{tabular}          & \begin{tabular}[c]{@{}l@{}}Payload\\Ground\end{tabular}                        & \checkmark                                    & \checkmark                                                                                                & i                                                                                                 & \checkmark                                                                                              & i                                                                                              & i                                                                                                   & i                                                                                              & x                                                                                                 & x                                                                                               \\ 
				\hline
				\begin{tabular}[c]{@{}l@{}}Bespoke\\Malware\end{tabular}                     & \begin{tabular}[c]{@{}l@{}}PLC Servo\\Exploit\end{tabular}                & \begin{tabular}[c]{@{}l@{}}Payload\\Ground\end{tabular}                        & \checkmark                                    & \checkmark                                                                                                & i                                                                                                 & \checkmark                                                                                              & \checkmark                                                                                            & \checkmark                                                                                                 & i                                                                                              & i                                                                                                 & x                                                                                               \\ 
				\hline
				\begin{tabular}[c]{@{}l@{}}Privilege\\Escalation\end{tabular}                 & \begin{tabular}[c]{@{}l@{}}Spotbeam\\Redirection\end{tabular}             & Payload                                                                        & \checkmark                                    & c                                                                                                  & \checkmark                                                                                               & x                                                                                                & \checkmark                                                                                            & i                                                                                                   & i                                                                                              & i                                                                                                 & x                                                                                               \\ 
				\hline
				Hijacking                                                                     & \begin{tabular}[c]{@{}l@{}}TT\&C Auth.\\Overwrite\end{tabular}            & Payload                                                                        & \checkmark                                    & c                                                                                                  & c                                                                                                 & x                                                                                                & c                                                                                              & i                                                                                                   & i                                                                                              & i                                                                                                 & x                                                                                               \\ 
				\hline
				\begin{tabular}[c]{@{}l@{}}Sensor\\Injection\end{tabular}                     & \begin{tabular}[c]{@{}l@{}}Falsified IR\\Signature\end{tabular}           & Payload                                                                        & \checkmark                                    & c                                                                                                  & x                                                                                                 & x                                                                                                & x                                                                                              & c                                                                                                   & x                                                                                              & x                                                                                                 & x                                                                                               \\ 
				\hline
				Jamming                                                                       & \begin{tabular}[c]{@{}l@{}}Broadcast\\Interruption\end{tabular}           & Signal                                                                         & \checkmark                                    & c                                                                                                  & x                                                                                                 & x                                                                                                & i                                                                                              & c                                                                                                   & \checkmark                                                                                            & i                                                                                                 & i                                                                                               \\ 
				\hline
				Eavesdropping                                                                 & \begin{tabular}[c]{@{}l@{}}IP Traffic\\Intercept\end{tabular}             & Signal                                                                         & c                                      & \checkmark                                                                                                & c                                                                                                 & c                                                                                                & \checkmark                                                                                            & c                                                                                                   & c                                                                                              & \checkmark                                                                                               & c                                                                                               \\ 
				\hline
				\begin{tabular}[c]{@{}l@{}}Metadata\\Analysis\end{tabular}                    & \begin{tabular}[c]{@{}l@{}}IP Traffic\\Fingerprinting\end{tabular}        & Signal                                                                         & c                                      & \checkmark                                                                                                & c                                                                                                 & x                                                                                                & i                                                                                              & c                                                                                                   & i                                                                                              & i                                                                                                 & x                                                                                               \\ 
				\hline
				\begin{tabular}[c]{@{}l@{}}Command\\Injection\end{tabular}                    & \begin{tabular}[c]{@{}l@{}}TT\&C\\Spoofing\end{tabular}                   & Signal                                                                         & \checkmark                                    & c                                                                                                  & \checkmark                                                                                               & x                                                                                                & \checkmark                                                                                            & i                                                                                                   & i                                                                                              & i                                                                                                 & x                                                                                               \\ 
				\hline
				\begin{tabular}[c]{@{}l@{}}Replay\\Attacks\end{tabular}                       & \begin{tabular}[c]{@{}l@{}}TT\&C\\Replay\end{tabular}                     & Signal                                                                         & \checkmark                                    & c                                                                                                  & \checkmark                                                                                               & x                                                                                                & \checkmark                                                                                            & \checkmark                                                                                                 & \checkmark                                                                                            & i                                                                                                 & x                                                                                               \\ 
				\hline
				\begin{tabular}[c]{@{}l@{}}Signal \\Injection\end{tabular}                    & \begin{tabular}[c]{@{}l@{}}Broadcast\\Piracy\end{tabular}                 & Signal                                                                         & c                                      & c                                                                                                  & \checkmark                                                                                               & x                                                                                                & c                                                                                              & c                                                                                                   & \checkmark                                                                                            & \checkmark                                                                                               & \checkmark                                                                                             \\ 
				\hline
				\begin{tabular}[c]{@{}l@{}}Generic\\Malware\end{tabular}                      & \begin{tabular}[c]{@{}l@{}}Windows\\Ransomware\end{tabular}               & Ground                                                                         & \checkmark                                    & \checkmark                                                                                                & i                                                                                                 & \checkmark                                                                                              & \checkmark                                                                                            & \checkmark                                                                                                 & \checkmark                                                                                            & \checkmark                                                                                               & \checkmark                                                                                             \\ 
				\hline
				\begin{tabular}[c]{@{}l@{}}Social\\Engineering\end{tabular}                   & \begin{tabular}[c]{@{}l@{}}Technology\\Theft\end{tabular}                 & Ground                                                                         & \checkmark                                    & \checkmark                                                                                                & \checkmark                                                                                               & c                                                                                                & \checkmark                                                                                            & \checkmark                                                                                                 & \checkmark                                                                                            & \checkmark                                                                                               & \checkmark                                                                                             \\ 
				\hline
				\begin{tabular}[c]{@{}l@{}}Physical\\Access\end{tabular}                      & \begin{tabular}[c]{@{}l@{}}Cleanroom\\Breach\end{tabular}                 & Ground                                                                         & \checkmark                                    & \checkmark                                                                                                & \checkmark                                                                                               & x                                                                                                & i                                                                                              & x                                                                                                   & i                                                                                              & i                                                                                                 & x                                                                                               \\ 
				\hline
				\begin{tabular}[c]{@{}l@{}}Data\\Corruption\end{tabular}                      & \begin{tabular}[c]{@{}l@{}}IMINT\\Corruption\end{tabular}                 & Ground                                                                         & \checkmark                                    & c                                                                                                  & \checkmark                                                                                               & x                                                                                                & \checkmark                                                                                            & i                                                                                                   & x                                                                                              & i                                                                                                 & x                                                                                               \\
				\bottomrule
			\end{tabular}
			\begin{tablenotes}
				\item Key:~\textbf{\checkmark}~-~Attacker is likely both capable of executing the attack and motivated to do so. ~\textbf{c}~-~Attacker is likely capable, but the vulnerability doesn't align with motivations. \textbf{i}~-~Attacker is likely interested in the attack, but has limited capacity to execute it. \textbf{x}~-~Attacker is likely neither interested in nor capable of executing the attack.
				\item Note: There may be crossover between categories, such as an insider threat sponsored by an intelligence agency. This matrix is intended as a demonstrative summary of likely outcomes, not a rigid proscription of all possible attacker motives and means.
			\end{tablenotes}
		\end{threeparttable}
	}
\end{table*}

\subsection{Unique Technical Security Challenges}
\label{sec:security-challenges}
A superficial reading of these vulnerabilities may suggest that satellites pose few novel challenges for systems security researchers. After all, terrestrial instances of all the listed vulnerabilities can easily come to mind. Indeed, many researches - especially from the commercial space sector - contend that traditional IT security approaches offer sufficient coverage, advocating for the use of NIST controls and generic security information and event management (SIEM) tools~\cite{knezLessonsLearnedApplying2016,youngCommercialSatellitesCritical2017,veraCyberSecurityAwareness2016,viveroCyberSituationalAwareness2018}. Beyond technical intuition, this viewpoint is commercially appealing as it allows for the direct use of widely available security tools (and cross-domain hire of experts in those tools) as the main line of defense for satellite missions~\cite{csricCybersecurityRiskManagement2015}.

As tempting as this viewpoint may be, it is not without detractors. Byrne et al., speaking primarily from the perspective of aerospace academia, argue that ``the assertion that existing controls will protect against risk is sometimes accepted without reasonable supporting data or, even worse, is accepted where the lack of data is used as proof''~\cite{byrneCyberDefenseSpacebased2014}. Falco, a computer science academic, takes this further, arguing that attempts to map traditional IT security to the space domain has created harmful technical knowledge gaps and discouraged specialization~\cite{falcoVacuumSpaceCyber2018}.

Falco further isolates six reasons that satellite cyber-security requires unique technical perspectives unmet by status-quo security practice~\cite{falcoVacuumSpaceCyber2018}. First, satellites represent a single point of failure for other critical infrastructures, increasing the number and capabilities of attackers who may be interested in harming them beyond that obviously relevant to mission function. Second, there is little regulation guiding satellite cyber-security, creating uncertainty regarding the controls appropriate to a given system. Third, complicated supply chains not only give rise to backdoor risks, but also make it difficult to assign organizational responsibility for security practice. Fourth, the widespread use of COTS hardware integrated with bespoke systems creates a unique situation where vulnerabilities likely apply to many platforms, but applying patches may require bespoke modifications. Fifth, the specialized nature of aerospace means that few individuals in cyber-security understand satellites sufficiently to adequately contextualize threats and defense. Finally, satellites are compute-constrained devices with limited resources and security/performance trade-offs are more acute than in terrestrial systems.

The second point, regarding the shortcoming of existing regulatory standards is further supported by Fidler, writing for the Council on Foreign Relations - an international relations policy think-tank~\cite{fidlerCybersecurityNewEra2018}. In particular, he contends that mappings of IT standards to space systems amount to little more than ``paper-shuffling''~\cite{fidlerCybersecurityNewEra2018}. Bardin contends that industry is unsure what would even constitute a cyber-attack against space systems, due to lack of comprehensive threat modeling~\cite{bardinSatelliteCyberAttack2013}. This may be attributable in part to overuse of the term ``hacking'' in media and policy circles to describe any disruption to satellite operations~\cite{fritzSatelliteHackingGuide2013,bardinSatelliteCyberAttack2013}. For example, technical authors often treat radio jamming as an unrelated topic while policy analysts explicitly consider it a cyber-attack vector~\cite{harrisonSpaceThreatAssessment2018,dscSpaceFinalFrontier2016,delmonteCybersecurityPolicySustainable2013}.

Falco's third and fourth point, regarding supply chains, have been subject to much attention as well. Space missions have uniquely complex bureaucratic structures. Many distinct organizations may share some device resources (e.g. communications systems), while operating others independently (e.g. on-board sensors). Excepting the largest players, satellite operators do not control the entire mission lifecycle. Launch vehicles, orbital injection, operation, and retirement are frequently handled by distinct entities. Some service providers (e.g. satellite television services) may have no ownership stake in the space platform at all, but instead simply lease radio access. The result is that operators of an information cannot necessarily trust each other and may not share security priorities. Any given member of the mission ecosystem can potentially compromise others~\cite{delmonteCybersecurityPolicySustainable2013}. This threat is particularly acute for ``New Space'' systems which rely heavily on third-party COTS equipment~\cite{laneHighAssuranceCyberSpace2017,veraCyberSecurityAwareness2016}.

Finally, Falco's ``expertise vacuum'' is widely recognized as a significant barrier. Niche components of satellite systems lack direct terrestrial equivalents (e.g. star-trackers), impairing the development of a general body of knowledge for securing these devices~\cite{veraCyberSecurityAwareness2016}. In academic contexts, the cross-disciplinary mixture of engineering, astrophysics, computer science, and security studies complicates the search for appropriate venues and communities for publication and peer-review. For example, expertise in cryptography may not be directly useful without additional hardware and astrophysics knowledge as extra-terrestrial radiation can induce random bit-flips in cryptographic key storage and requires special attention~\cite{banuOnBoardEncryptionEarth2006}.

Ultimately, space systems are much more than mere ``computers in the sky.'' Well-regarded terrestrial security practices often fail to transfer to space systems for unintuitive reasons which require a wide breadth of expertise to overcome. The result is that relatively little work, especially within systems security, has been conducted on space technologies.

\section{Learning from History}
Given the dearth of academic satellite cyber-security research, the threat may appear distant and hypothetical. Indeed, few satellite hacking incidents over the past half-century have received significant public attention and one might be tempted to argue that satellite cyber-security is more an invented problem than present danger.

\begin{figure}[h]
	\includesvg[width=\linewidth, inkscapelatex=false]{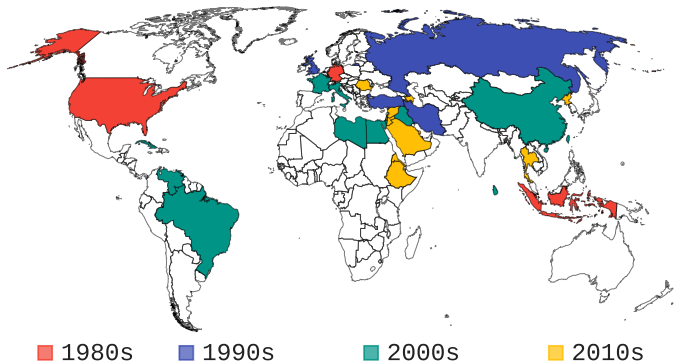}
	\label{fig:world-map}
	\caption{Countries Involved in Satellite Hacking by Year of First Entry. This chart includes countries with governments and/or citizens implicated as attackers in satellite security incidents.}
\end{figure}

However, a deeper look at the history of operations targeting satellites reveals an unconventional but voluminous body of knowledge. Indeed, cyber attacks against satellite systems have been occurring, almost unnoticed, for decades - perpetrated by attackers from across the globe (Figure~\ref{fig:world-map}). In this section, we present an overview of this empirical data with a focus on long-term trends and unsolved security problems. This analysis builds on the prior work of Fritz and Manulis et al.~\cite{fritzSatelliteHackingGuide2013,manulisCyberSecurityNew2020}. 

In conjunction with this paper, we have developed an annotated chronology which details 113 significant satellite hacking incidents from 1957 to present day (Appendix~A). To our knowledge, this chronology represents the most exhaustive record of satellite hacking incidents to date. Derived from original archival research synthesizing unclassified primary and secondary source materials, it offers evidence-based technical insights into the evolution and practice of satellite exploitation.

Before delving into the survey, it is worth clarifying its scope. In particular, the topicality of RF interference has been subject to much debate - with some regarding it as an issue of electronic warfare as opposed to cyber-operations. We have elected to include some of the most notable instances of such attacks in our analysis for two reasons. First, a willingness to engage in jamming suggests that an attacker values the ability to ``virtually'' deny satellite access to their victims - offering potential insights into threat models for digital counterspace. Second, the hardware and expertise involved jamming operations often has significant cross-over with more obviously topical signal-hijacking and injection attacks.

\subsection{1957-1979: Early Days}
In the earliest days of human spaceflight, the principle information security concerns revolved around the ability of adversaries to compromise satellite flight control signals. One of the first public discussions of satellite information security was a 1962 US congressional hearing to determine if private companies should be allowed to operate in space~\cite{newyorktimesRadioSatellitesOpen1962}. It was suggested that commercial missions would be more vulnerable to jamming and replay attacks from Soviet adversaries, while higher-altitude military satellites were presumed secure due to the complexity of the requisite equipment.

The subsequent two decades saw no major satellite hacking incidents. However, a important political debate was brewing over satellite broadcast abuse. The US had begun transmitting anti-communist propaganda on satellite beams directed into Soviet territory. In response, the USSR put forward a UN proposal in 1972 asserting a sovereign right to jam illegal radio signals in their territory~\cite{washingtonpostSovietsAskControls1972}. To this day, state sovereignty over radio emanations from foreign satellites remains contentious. Modern norms on interstate jamming and eavesdropping attacks can be readily traced back to this 1972 dispute.

\subsection{1980-1989: Piracy and Spoofing}
The first major satellite hacking incident is generally thought to have occurred in 1986. An industry insider and satellite-dish salesman pseudonymously dubbed ``Captain Midnight'' hijacked an HBO television broadcast destined for satellite TV customers in Florida and replaced it with a message chastising network executives for new signal-scrambling copy protection technology~\cite{davisCaptainMidnightUnmasked1986}. Interestingly, this attack almost exactly mirrored a fictional short-story from a satellite enthusiast magazine the previous year - although no formal association has been proven~\cite{pollackTechnology1986}. The next year, a similar attack took place wherein an employee of the Christian Broadcasting Network replaced a satellite stream operated by The Playboy Channel with biblical verses chastising viewers for not attending church on Sunday~\cite{smithAppealUnitedStates1991}.

1986 also marked the first major satellite eavesdropping case, wherein the government of Indonesia was accused by an American satellite imaging company of illegally intercepting earth observation data without paying for a subscription to the satellite's service~\cite{timesofindiaSpacePiracy1986}.

Terrestrially, the 1980s marked the first major attack against satellite ground systems. In 1987, a group of West German teens compromised top secret NASA networks by means of a Trojan Horse program which concealed a keylogger~\cite{parryYouthsHackedSecret1987}. These networks were reported to include information on classified military space missions and to have the capability to cause direct harm to satellites. Upon intercepting a mail-box message indicating that the compromise had been discovered, the teenagers voluntarily turned themselves in.

\subsection{1990-1999: Broadcast and Flight Control Systems}
Both satellites usage and exploitation accelerated throughout the 1990s. As satellite television became commonplace, states began to use jamming attacks to control the flow of information across their borders. Iran began jamming foreign satellite television stations in 1994, a practice which continues today~\cite{afpForeignSatellitePrograms1994,smallmediaSatelliteJammingIran2012}. In 1998, Indonesia became the first country to deliberately use a satellite to jam signals from a neighboring satellite as part of a dispute with Hong Kong over orbital slot access~\cite{wongMilitarySpacePower2010, fritzSatelliteHackingGuide2013}. By the end of the 1990s, commercially available satellite jammers emerged on the market, including a \$4,000USD Russian-made device capable of disabling GPS signals over a 200~km radius~\cite{intelligencenewsletterAnybodyNeedGPS1998}.

The 1990s saw the widespread emergence of cryptographic systems for satellite television piracy - kicking off an ongoing battle between satellite pirates and media companies which began with simple smart-code sharing networks and escalated into sophisticated cryptanalysis~\cite{hellenHackersTargetBskyB1993, walshCableTVPirate1994}. From 1993 onwards, reports detail an essentially annual cycle of hackers breaking TV protections, media companies designing improvements, and governments making related arrests.

Finally, a number of attacks against satellite ground stations occurred over the 1990s. These included high-profile incidents where hackers claimed to have accessed to systems which would allow to issue flight control commands to orbiting satellites. Most notable among these is a 1998 scenario wherein hackers, widely believed to be Russian-government affiliated, gained access to flight control systems in NASA's Goddard Space Flight Center~\cite{elginNetworkSecurityBreaches2008,fritzSatelliteHackingGuide2013}. During this incident, the German-US ROSAT x-ray telescope inexplicably altered its orientation to point optical sensors directly at the sun - leading to irreparable hardware damage~\cite{elginNetworkSecurityBreaches2008, fritzSatelliteHackingGuide2013}. Although details surrounding the incidents are highly classified, this is often cited as the first cyber-attack which caused physical damage in orbit.

\subsection{2000-2009: Organized Attackers}
The 2000s saw more incidents than the previous forty years combined. One major trend was the emergence of organized non-state attackers. Notable incidents of this nature included signal hijacking attacks by Falun Gong (a Chinese religious and protest movement) from 2002-2005, similar attacks by the Tamil Tigers (a Sri Lankan militant organization) from 2007-2009, and eavesdropping attacks compromising US military drone video feeds by Iraqi insurgents in 2009~\cite{associatedpressFalunGongHijacks2002,southchinamorningpostFalunGongAccused2003,xinhuaAsiaSatAccusesFalungong2004,dalyLTTETechnologicallyInnovative2007,gormanInsurgentsHackDrones2009}.

Government-led jamming operations continued unabated. Most notable among these were an instance of Iranian jamming of signals directed to Turkey in 2000 and Cuban jamming of signals destined for the Middle East in 2003~\cite{afpIranianGovernmentJamming2000,marquezUSCondemnsCuba2003}.

Significant attacks against groundstations during this period include complete flight control takeover of two NASA satellites in 2007 and 2008~\cite{arthurChineseHackersSuspected2011,bardinSatelliteCyberAttack2013,zattiProtectionSpaceMissions2017}. These attacks were originally reported as signal jamming but later linked to a Chinese government compromise of NASA ground stations~\cite{arthurChineseHackersSuspected2011}.

The 2000s also saw the first public case of a malware infection in orbit. In 2008, a Russian cosmonaut introduced Windows-XP malware to systems aboard the International Space Station (ISS). This incident is widely believed to have been accidental~\cite{francisComputerVirusInfects2008, gibbsInternationalSpaceStation2013, zattiProtectionSpaceMissions2017}.

Although not directly related to cyber-security, a major space security incident occurred in January of 2007 when China demonstrated an anti-satellite (ASAT) weapon~\cite{zissisChinaAntiSatelliteTest2007}. Not only did this generate a significant amount of space debris, it also demonstrated emerging state interest in offensive counterspace technology. This ASAT demonstration was preceded by a less well known ``virtual'' attack in 2006, when a Chinese ground-based laser system was used to blind sensors aboard a classified US military satellite~\cite{couriermailChinaTargetsUS2006}.

\subsection{2010-Present: Evolving Threats}
The accelerating usage of cyber-operations in space has continued over most recent decade. In particular, a wave of jamming incidents in the Middle East and North Africa were kicked off by the Arab Spring protest movements in 2010 and have continued thereafter. This caused the list of countries with demonstrated satellite jamming capabilities to more than double with the addition of Egypt, Jordan, Bahrain, Ethiopia, Saudi Arabia, Eritrea, Syria, Azerbaijan, and Israel - along with renewed jamming from Libya and Iran~\cite{bbcmonitoringworldmediaUKbasedAlHiwarSatellite2009, bbcEUPressuresIran2010,bbcmonitoringmiddleeastJammingAlJazeeraTV2010,messiehBahrainSatelliteChannel2011,ecadfChinaAccusedJamming2011, thurayapressofficeThurayaTelecomServices2011,bbcmonitoringworldmediaIranArabicTV2011,leoHerosOrdinaires2013,richardsonEritreaAccusesEthiopia2012,bbcmonitoringworldmediaWorldBroadcastersCondemn2012, bbcmonitoringtranscaucasusunitAzeriEditorSays2013, bbcmonitoringmiddleeastSyrianTVStill2018}. Outside of the region, North Korea also began a sustained jamming campaign against South Korean military GPS in 2010~\cite{bbcmonitoringasiapacificNorthKoreaIncreases2013}.

More sophisticated signal-related attacks also emerged. This included signal intrusion attacks by Hamas against Israeli news stations and academic research demonstrating weaknesses in satellite internet, messaging, and telephone services~\cite{bbcmonitoringworldmediaHamasHacksSatellite2014,housen-courielWhenHamasComes2016,egeaPlayingSatelliteEnvironment2010,mooreSpreadSpectrumSatcom2015,driessenDonTrustSatellite2012,cryptoeprint:2017:655,pavurSecretsSkyPrivacy2019,pavurTaleSeaSky2020}. In 2014, Russia was accused of launching a ``stalker sat'' which followed other satellites in orbit to intercept uplink signals, representing the first publicly acknowledged instance of satellite-to-satellite eavesdropping~\cite{insidesatellitetvRussiaEavesdroppingSatellite2015}.

Attacks against ground-stations and satellite control systems grew more sophisticated as well, with many being linked to state actors. In particular, China has been accused of compromising US space control systems in 2011, 2014, and 2017~\cite{symantecThripEspionageGroup2017,flahertyChineseHackWeather2014,newmanReportAnotherChinese2014,bbcHackersControlledNasa2012}. This is perhaps unsurprising given that, in 2014, an internal US audit of the Joint Polar Satellite System (JPSS) ground stations found more than 9,000 ``high-risk'' security issues, many of which remained unpatched from prior audits~\cite{crawleyExpeditedEffortsNeeded2014}. Commercial ground systems were also demonstrated to have severe vulnerabilities, including many hardcoded passwords and backdoors~\cite{santamartaLastCallSATCOM2018,santamartaSATCOMTerminalsHacking2014}.

This period has also seen the first organized criminal abuse of satellite systems. In 2016, the Russian advanced persistent threat (APT) actor dubbed ``Turla group'' was found to be abusing satellite internet signals to anonymously exfiltrate data from compromised computer systems~\cite{tanaseSatelliteTurlaAPT2015}. This exfiltration method was further detailed by a security researcher at the DEFCON conference in 2020~\cite{pavurWhispersStars2020}.

In recent years, new attention has been paid to the satellite cyber-security field. In 2020, the US Air Force hosted an online ``Hack-A-Sat'' competition which explicitly sought to introduce cyber-security professionals to the world of satellite cyber-security and to uncover vulnerabilities in real space systems~\cite{defensedigitalserviceHackasat2020}. Similarly, in 2020, DEFCON hosted its first ``aerospace village,'' a sub-conference which included a briefings track focused exclusively on space systems security~\cite{aerospacevillageAeroSpaceVillageSecuring2020}.

\subsection{General Trends and Developments}
\begin{figure}[h]
	\label{fig:attack-types-history}
	\includesvg[width=\linewidth, inkscapelatex=false]{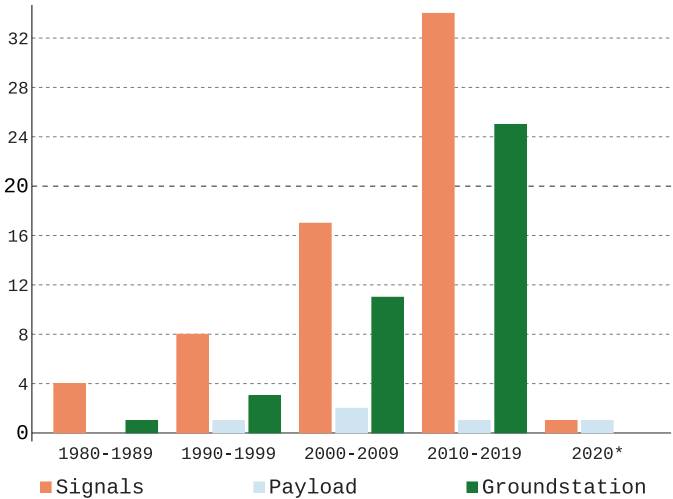}
	\caption{Satellite Attack Types by Decade. Data from 2020 is included for completeness on the basis of publicly available reports as of August 15, 2020. In practice, there is often substantial lag between intrusions, detection, and reporting.}
	
\end{figure}
In sum, there has been a clear general trend towards increased use of cyber-capabilities that target satellite systems~(Figure~\ref{fig:attack-types-history}). Over the past 60 years, and especially over the past 20, the number of actors willing and able to attack satellites in cyberspace has increased dramatically.

\begin{figure}[h]
	\label{fig:attacker-types-history}
	\includesvg[width=\linewidth, inkscapelatex=false]{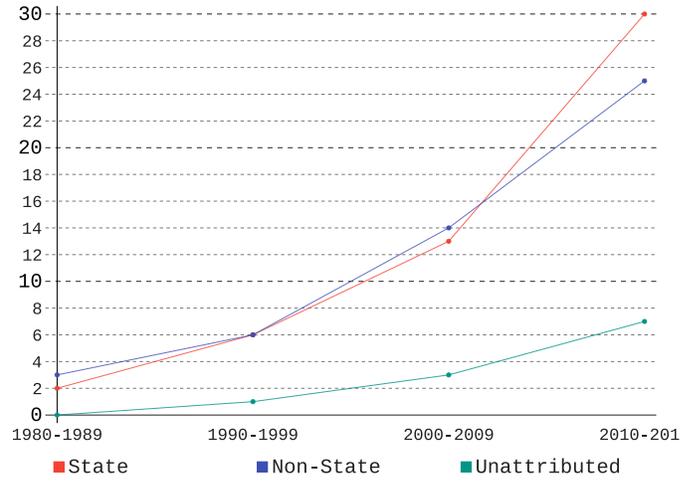}
	\caption{Satellite Attacks Associated With State and Non-State Actors. NB: Attribution is often uncertain and subject to dispute. Further detail regarding the attribution of any particular incident can be found in the open-source dataset referenced in Appendix~A.}
	
\end{figure}

Today, almost 30 states have demonstrated some degree of cyber-offensive counterspace capabilities, including many which lack spacefaring capabilities. Moreover, there has been a distinct rise in the frequency, complexity, and magnitude of attacks instigated by non-state actors~(Figure~\ref{fig:attacker-types-history}). Contrary to common perception, there is little historical evidence indicating that non-state actors are less willing or able to engage in digital-counterspace than state counterparts. However, this may be due to a reporting bias whereby non-state incidents are widely covered but nation-state attacks are classified.

Together, these trends clarify the need for research combating cyber-security threats to satellites. Attacks against satellites are happening in the status quo and have been for decades. As attackers grow more sophisticated and prevalent, increased awareness of present practice is a key first-step towards contributing meaningful technical research.

The remainder of this systematization of knowledge paper delves deeper into this chronology to identify unsolved technical questions in satellite cyber-security. These historical incidents are contextualized vis-a-vis the vulnerability matrix outlined in Section~\ref{sec:security-challenges} and organized on the basis of technical subsystems (RF, Space, Ground, and Mission).

\section{Defending The Signal}
More than two thirds of historical satellite incidents in our review related to attacks on the RF communications link.

A significant portion of these are best classified as ``jamming'' attacks, which tend to require physical mitigations - such as frequency hopping. As our focus is on digital counterspace, as opposed to electronic warfare, we will not delve deeply into jamming, but it is worth noting that the anti-jamming field is well-developed~\cite{simoneFrequencyHoppingTechniquesSecure2006, gunnAnomalyDetectionSatellite2018}. Further, jamming incidents often demonstrate important political context regarding attacker motivations and equipment capabilities.

Beyond jamming, we outline three general categories of communications attacks in our historical analysis. The first, eavesdropping, relates to the interception and interpretation of signals by an unintended third-party recipient. The second, signal injection, relates to the encapsulation of malicious data inside an otherwise legitimate radio transmission. The final category, signal spoofing, relates to attempts to artificially hijack and replace legitimate radio signals with malicious ones.

\subsection{Eavesdropping Attacks}
\label{sec:eavesdropping}

The eavesdropping challenge for satellites is primarily one of scale. Signals from a single geostationary (GEO) satellite can encompass an entire continent due to the vast transmission distances. This means that attackers across a wide range of jurisdictions may be capable of receive the transmission and that sending sensitive data unencrypted over such signals is generally ill-advised. In our historical review, we find that the scope and frequency of eavesdropping incidents has increased significantly overtime. It has been suggested that this is largely due to widespread access to the requisite equipment - such as Software Defined Radios (SDRs) - at reduced costs~\cite{weinbaumSigintAnyoneGrowing2017, pavurTaleSeaSky2020}.

Despite a clear case for the use of encryption in satellite environments, the practical implementation of satellite crypto-systems is quite complex. Satellite signals travel over immense distances and are frequently subjected to significant packet loses and high latency due to speed-of-light constraints~\cite{roy-chowdhurySecurityIssuesHybrid2005}. The naive addition of terrestrial encryption schemes to satellite environments can have severe negative impacts on overall performance. By some estimates, this can amount to as much as 80\% reduction in perceived performance~\cite{fritzSatelliteHackingGuide2013, pavurQPEPQUICBasedApproach2020}.

Most satellite encryption techniques focus on ground-based encryption, treating the satellite as a ``bent-pipe'' for signal relay. However, in cases such as the transmission of Telemetry, Tracking and Command (TT\&C) data, on-board encryption capabilities may be required. Satellite hardware is heavily resourced constrained and subject to a harsh orbital environment and thus on-board encryption is a non-trivial challenge~\cite{banuOnBoardEncryptionEarth2006}. Indeed, improperly implemented on-board encryption may be perverted by attackers into a denial-of-service mechanism by sending large quantities of deliberately invalid data to overwhelm limited computational capabilities~\cite{roy-chowdhurySecurityIssuesHybrid2005}.

\subsubsection{Encryption in Broadcast Networks}
In broadcast satellite environments, such as those used for television and radio services, a number of cryptographic solutions have emerged. This development has been driven by a commercial need to restrict satellite television access to paying customers. However, our historical review outlines a perpetual ``game of cat-and-mouse'' between satellite TV operators and attackers.

One of the most widely used systems in this context is the Common Scrambling Algorithm (CSA), which encrypts Digital Video Broadcasting (DVB) streams with a hybrid combination of stream and block ciphers~\cite{wirtFaultAttackDVB2005}. CSA has been found to have severe weakness which make it possible to crack most streams in real-time on consumer hardware~\cite{liSecurityAnalysisDVB2007,wirtFaultAttackDVB2005,tewsBreakingDVBCSA2011}.

Alternative schemes are often proprietary and based on the use of smart-cards or specialized receivers with pre-distributed keys. An example is the DigiCipher format which accounted for around 70\% of encrypted satellite broadcasts in North America in 2012~\cite{bardinSatelliteCyberAttack2013}. Another popular system is the PowerVu, which is used by the American Forces Network~\cite{bardinSatelliteCyberAttack2013}. In 2014, it was demonstrated that PowerVu root management key entropy could be trivially reduced to a 16~bits, enabling real-time attacks on the system~\cite{colibriPowerVuManagementKeys2014}.

In general though, most attacks have targeted key distribution rather than cryptography. Smart cards, for example, are often emulated or copied to share one legitimate subscription among hundreds of illegitimate users. This works because broadcast signals are often encrypted with a single key which all customers must be capable of deriving. France et al. proposed a process by which individual keys could be revoked without re-issuing cards to all legitimate customers~\cite{francisCountermeasuresAttacksSatellite2005}. In general, academic work on the topic has focused on this key-revocation problem, but enduring solutions have proven elusive~\cite{howarthDynamicsKeyManagement2004,roy-chowdhurySecurityIssuesHybrid2005,shengSecurityArchitectureSatellite2011,hughesQuantumCryptographySecure2000,sesSESAnnounces102018}.

\subsubsection{Encryption in IP Networks}
For internet and broadband services, encryption is more complex. Due to speed-of-light latency, particularly in long-range GEO networks, TCP can suffer several negative performance effects~\cite{roy-chowdhurySecurityIssuesHybrid2005,fidlerSatelliteNewOpportunity2002}. Satellite ISPs mitigate these issues and preserve limited bandwidth through the use of active traffic manipulation~\cite{iyengarSecurityRequirementsIP2007,roy-chowdhurySecurityIssuesHybrid2005}. This requires ISPs to have direct access to customer TCP headers and, in some cases, HTTP payloads. As a result, the use of VPNs and customer-implemented end-to-end encryption results in significant performance reductions.

Several solutions have been proposed to protect traffic over-the-air while maintaining acceptable performance. For example, Roy-Chowdhury et al. suggests the use of a multi-step SSL variant reveals certain header information to ISPs while leaving payload data encrypted~\cite{roy-chowdhurySecurityIssuesHybrid2005}. Duquerroy et al. proposed a modification of IPSec called SatIPsec which provides a layer-three encrypted tunnel with support for multicasting encryption~\cite{duquerroySatiPSecOptimizedSolution2004,iyengarSecurityRequirementsIP2007}. However, this solution also granted ISPs access to some customer traffic and required pre-shared secrets~\cite{roy-chowdhurySecurityIssuesHybrid2005}. More recently, Pavur et al. have developed an open-source proxy which leverages the UDP-based QUIC protocol for over-the-air encryption and is intended for customers to deploy on their personal devices~\cite{pavurQPEPQUICBasedApproach2020}. 

In practice, many satellites ISPs use none of these solutions, instead sending sensitive customer traffic in clear-text. This has been shown to impact the security and privacy of home internet customers, critical infrastructure systems, maritime vessels, and aviation networks~\cite{pavurWhispersStarsPractical2020}.

\subsection{Signal Injection Attacks}
\begin{figure}[h]
	\includesvg[width=\linewidth, inkscapelatex=false]{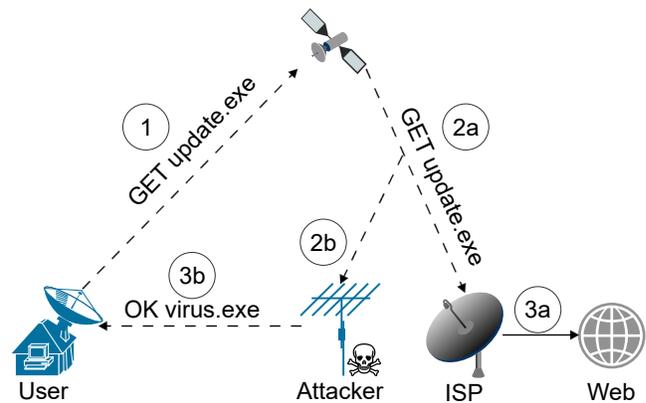}
	\caption{Demonstrative Signal Injection Attack. At time 1 in the figure, the user requests a secure download file from a trusted server. Both the ISP (2a) and the eavesdropper (2b) receive this transmission concurrently. While the ISP routes the request over the internet (3a), the attacker transmits a malicious response to the customer antenna (3b).}
\end{figure}
Significantly less research has been conducted on signal injection attacks. Historically, satellite companies operated under the assumption that the complexity and cost of requisite equipment to alter or misuse legitimate satellite signals was beyond the means of most attackers~\cite{newyorktimesRadioSatellitesOpen1962}. However, novel attacks requiring little to no specialized equipment appeared in our historical analysis. For example, the Turla group attacks uncovered by Kaspersky in 2015 demonstrated that simply transmitting normal web-requests to IP addresses in a satellite ISP's network would result in those messages being injected into satellite broadcasts~\cite{tanaseSatelliteTurlaAPT2015,pavurWhispersStars2020}. Similarly, a security researcher demonstrated that software defined radios were sufficient to transmit specially crafted packets on the Globstar network, despite their use of complex Distributed Spread Spectrum (DSS) signaling~\cite{mooreSpreadSpectrumSatcom2015}. Further, the theoretical threat has been suggested, but not demonstrated, that an attacker could inject packets directly into a user's receiving antenna by emulating a satellite - allowing them to compromise programs running on the victim's machine and bypass many firewall restrictions~\cite{iyengarSecurityRequirementsIP2007}. Lane et al. argues that carefully crafted packets may even be used to trigger vulnerabilities in the networking hardware stacks on the satellite itself~\cite{laneHighAssuranceCyberSpace2017}.

To the best of our knowledge, no general defense against signal injection has been proposed. However, many of the encryption protocols discussed in section~\ref{sec:eavesdropping} may also bolster the general integrity of satellite signals. Additionally many TT\&C encryption standards, such as Space Data Link Security (SDLS), would intuitively complicate these attacks~\cite{ccsdsSpaceDataLink2018}.

\subsection{Spoofing}
The final category of signals-based attacks identified in our historical review is that of signal spoofing. The form and severity of these attacks has varied widely between incidents. However, the most common variant are attacks targeting media broadcasts - generally satellite television signals. Here, attackers typically replace the attacker's uplink signal with a more powerful malicious radio transmission~\cite{dscSpaceFinalFrontier2016}. 

As broadcast satellites often operate as dumb ``bent-pipes,'' they will dutifully relay any incoming transmission on the correct frequency. The most intuitive protection against such attacks would be on-board verification of incoming signals. To the extent that such mechanisms exist in the status quo, they rely on proprietary and bespoke trade secrets which have not been well studied. To the best of our knowledge, no public on-board verification standard for satellite broadcasts exists. Such a system is non-trivial to design due to compatibility requirements with legacy ground-stations, high cost of replacing orbital hardware, and general difficulties with encryption hardware in space (see Section~\ref{sec:eavesdropping}).

One variant of signal-spoofing attacks which has received substantial academic attention relates to the spoofing of Global Navigation Satellite System (GNSS) signals, such as those from the US-operated Global Positioning System (GPS). Because GNSS signals are quite faint by the time they reach Earth, attackers can overpower these transmissions locally using inexpensive and widely available equipment. GNSS spoofing has been studied since at least the late 1990s, but the recent emergence of consumer-grade HDR hardware has made it possible for even hobbyists to spoof GNSS signals~\cite{intelligencenewsletterAnybodyNeedGPS1998, Tippenhauer:2011:RSG:2046707.2046719}. Indeed, in 2016, SDR-enabled wireless GPS spoofing attacks were used by players of the popular mobile game Pokemon GO as a cheating mechanism~\cite{rtl-sdr.comCheatingPokemonGo2016}.

The simplest GNSS spoofing attacks target terrestrial systems and involve directing a false simulated GNSS signal towards the victim receiver~\cite{Humphreys08assessingthe}. More complicated attacks seek to avoid detection by, for example, correcting time synchronization discrepancies or modifying known valid GNSS signals rather than simulating them from scratch~\cite{Humphreys08assessingthe}. The most sophisticated attacks may go further, simulating the spatial distribution of the originating GNSS satellites to emulate expected physical signal characteristics~\cite{wesson2012straight}.

Dozens of defenses against GNSS spoofing attacks have been proposed. These range from sanity checking GNSS readings with additional sensor data (e.g. using an accelerometer to identify GNSS motion that does not correspond to physical motion) to spectrum anomaly detection to flag the presence of spoofed transmissions against a historic baseline~\cite{warner2003gps,wen2005countermeasures}. A full treatment of state-of-the-art research on GNSS counter-spoofing could easily exceed the length of this paper and is well beyond our prerogative. As a starting point, Jafania-Jahromi et al. provide an accessible but deep survey of more than a dozen different classes of GPS anti-spoofing techniques, including techniques which allow individuals to determine their location accurately in the presence of an attacker~\cite{jafarnia2012gps}. 

Beyond GNSS spoofing, little research attention has been paid to the spoofing of satellite broadcasts. These range from satellite internet services to specialized critical infrastructure communications links. Given the relative maturity of the GNSS security community, it is possible lessons learned there may prove applicable to related challenges. Future research which considers the utility of GNSS counter-spoofing techniques for non-GNSS transmissions may be a promising approach to defending satellite signal authenticity.

\subsection{Key Concerns for Satellite Signals Security}
At a high level, the dominant security challenge for satellite RF links is their inherently public physical nature. While similar issues have been mitigated terrestrially (e.g. in cellular networks), the unique hardware and environmental constraints of orbit mean few terrestrial solutions are ``drop-in'' compatible with satellites. This has impeded the widespread adoption of link-layer encryption. Even when defenses are widely employed, such as in broadcast television, they often depend on proprietary ``black-box'' encryption which have repeatedly proven vulnerable to exploitation.

Without robust and open signal security protocols which consider the unique demands of space, attackers will continue engaging in sophisticated eavesdropping, injection, and spoofing attacks. This goes beyond the academic task of inventing crypo-systems. Many well-studied, such as SatIPSec, have been largely ignored due to their complexity and cost of adoption~\cite{duquerroySatiPSecOptimizedSolution2004}. Effective future work must incorporate not only technical systems-security perspectives, but also pragmatic understandings of the commercial and operational needs of satellite operators.

\section{Defending Space Platforms}
\label{sec:platform-security}
When compared with the RF domain, only a small amount of literature exists on the defense of satellite payloads themselves. This dearth of research likely results from a few factors. First, satellite payloads have historically been highly bespoke systems~\cite{falcoVacuumSpaceCyber2018,viveroSpaceMissionsCybersecurity2014,fritzSatelliteHackingGuide2013}. Academics seeking broadly novel scientific findings may struggle to generalize from issues relating to any specific platform. This is further compounded by the proprietary and often restricted nature of satellite hardware, with export controls impeding trans-national collaboration. Finally, the industry acts as a ``gatekeeper'' to many of these components and often demonstrates skepticism or even hostility towards security research~\cite{csricCybersecurityRiskManagement2015}. 

Also worth noting is that aerospace academia differs substantially from systems security academia. Even ``simple'' CubeSat projects are multi-year endeavors involving dozens or hundreds of collaborators~\cite{eoportalSwissCube}. A PhD thesis in aerospace engineering may revolve around the design of a single sub-system for such missions~\cite{pietzkaDevelopmentCharacterizationPropulsion2016}. Publications, especially prior to launch, tend to consist of narrow descriptions of engineering and implementation details, with less focus on broad theoretical generalizations and more focus on practical lessons and novel techniques. It is not unusual to encounter aerospace research describing the results of months or years of 3D-CAD modeling or simulation in less than five pages, with the model or satellite itself constituting the main contribution. 

The relative verbosity, fast publication rhythm, and paper-first culture of systems security academia complicates cross-domain collaboration. On some topics, such as RF communications, this matters little; security academics can manage financially and technically without leaning on aerospace counterparts. However, space platforms are inordinately complex and expensive. Security researchers wishing to ``go it alone'' can struggle to make meaningful headway. Future work which demonstrates interdisciplinary collaboration that fulfills the career and scientific objectives of participants from both fields is a much needed contribution - perhaps even irrespective of the substance of the research.

As satellite development undergoes significant changes, collaboration may become slightly easier. For example, the increased use of COTS components lowers to cost of new research collaborations with increasing the chance of broadly generalization findings.

Despite present day barriers, there has been a small but meaningful quantity of prior work on payload security. Wheeler et al. outlines four broad attack surfaces: input systems like sensors and RF antennae, output systems such as telemetry transmitters, internal communications such as Spacewire buses, and the underlying flight computer which integrates these components~\cite{wheelerCyberResilientFlight2017}. They further offer a ``top-ten'' list of possible vulnerabilities, ranging from malformed sensor data leading to buffer-overflows to malicious triggered safe-mode status caused by unhandled hardware states~\cite{wheelerCyberResilientFlight2017}.

The primary status quo defense against such attacks is boundary delineation. Through RF encryption and specialized groundstation hardware, satellite operators mitigate the risk of malicious individuals issuing instructions which could trigger unintended in-orbit behaviors.

\begin{figure*}
	\includesvg[width=\textwidth, inkscapelatex=false]{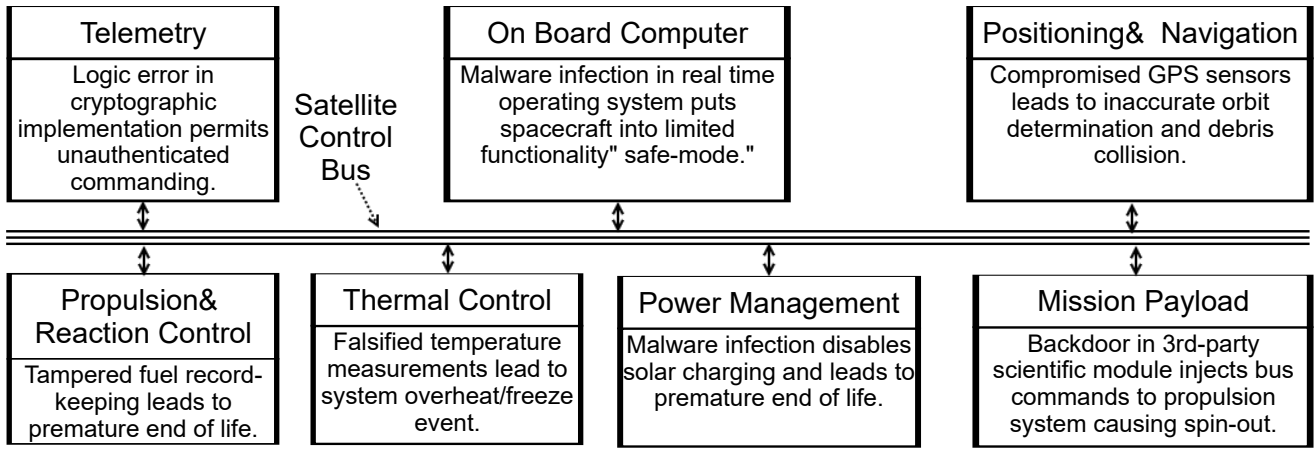}
	\caption{A simplified satellite architecture with example compromise scenarios for onboard sub-systems.}
	\label{fig:onboard-attacks}
\end{figure*}

Cohen et al. takes issue with this strategy, contending that it creates an ``open-trust'' environment in space~\cite{cohenSpacecraftEmbeddedCyber2016}. Once boundary protections are overcome, lateral movement aboard the spacecraft and privilege escalation are trivial. This increases the threat posed by backdoors introduced into the spacecraft during its development on the ground. For example~\ref{fig:onboard-attacks} provides a high-level schematic of on-board satellite sub-systems and scenarios where compromise of each could escalate to mission failure.

This problem is not easily mitigated as cyber-attacks and environmentally-induced hardware malfunction are often indistinguishable to a ground observer~\cite{wheelerCyberResilientFlight2017}. The remoteness of space means that forensic auditing capabilities must be built prior to launch and remain uncompromised following an attack~\cite{cohenSpacecraftEmbeddedCyber2016}. Moreover, the limited bandwidth, data-storage, and compute capabilities of satellites means that it is rarely economical to store or transmit complete audit logs~\cite{cohenSpacecraftEmbeddedCyber2016}. The transmission and storage of security data directly competes with the core mission functionalities.

To mitigate these issues, both Cohen et al. and Wheeler et al. independently suggest the adoption of an on-board monitoring agent which detects behavioral anomalies and engages in autonomous intrusion prevention~\cite{wheelerCyberResilientFlight2017,cohenSpacecraftEmbeddedCyber2016}. This would facilitate clearer auditing and recovery in response to malicious behaviors, but, if not implemented correctly, could trigger harmful false-positives. Unfortunately, this approach lacks backwards compatibility, although some basic functionality (such as audit logging) may be applied to existing satellites~\cite{wheelerCyberResilientFlight2017}.

In addition to a general security monitoring component, it has been suggested that satellite hijacking attempts could be prevented through frequent, automatic re-imaging of satellite software~\cite{llansoAchievingSpaceMission2016}. By storing a verified secure copy of the satellite operating system on a trusted platform module (TPM) it may be possible to bolster resilience by limiting the amount of time which an attacker might abuse the system. There are two notable downsides to this approach. The first is that it requires the addition of new hardware components (the TPM) - increasing satellite weight and power drain. The second is that it makes it difficult for satellite operators to securely patch to vulnerabilities as, in the case of a read-only firmware backup, they would be overwritten.

Some attention has also been paid to the flight code itself. Wheeler et al. notes that more than 95\% of the alerts raised by conventional code analysis tools triggered false positives on one demonstrative satellite, and they suspect many false negatives also occurred~\cite{wheelerCyberResilientFlight2017}. It has been suggested that formal verification may mitigate these issues, but no practical solution has been demonstrated to date~\cite{laneHighAssuranceCyberSpace2017}. Satellite software is also rarely monolithic, incorporating third-party code for various components and increasing the risk of software backdoors.

In sum, payload security is a critical but understudied topic. Prior work has demonstrated a wide range of severe and unmitigated attack vectors. While barriers to research are particularly acute, there is clear need for future technical work.

\section{Defending Satellite Ground Systems}
Unlike space platforms, which suffer from esoteric hardware and limited access, ground systems benefit from the wealth of general cyber-security knowledge. Typically, satellite ground stations are not distinct from any other terrestrial computing network and, where they do differ, remain similar to terrestrial communications systems~\cite{youngCommercialSatellitesCritical2017}. Although some diversity of implementation exists, all ground stations at a minimum consist of radio equipment to communicate with satellites and a computer/modem which operates this equipment. Normally, the computer will run traditional operating systems with specialized software for satellite communications.

On rare occasions, our historical review found this specialized software to be the target of attacks. For example, in 2000 hackers stole copies of Exigent satellite control software for the purpose of reverse engineering~\cite{lemosSatelliteControlCodes2001}. More typically, attacks are byproducts of general, untargeted, intrusions (e.g. in 1999 when a curious teenage hacker accidentally gained access to NASA flight control systems~\cite{wilsonTeenGivenSix2000}). Because of this, very little academic literature focuses on ground station security. Nevertheless, some unique aspects are worth consideration.

First, satellite ground systems almost always represent the final security boundary against payload exploitation~\cite{cohenSpacecraftEmbeddedCyber2016}. As discussed in Section~\ref{sec:platform-security}, satellite software and hardware typically follow an ``open trust'' model whereby the ground station is trusted by all devices aboard the space platform. As such, ground systems represent a single point of failure for satellite missions. In light of this problem, Llanso and Peaerson suggest the development of redundant restricted-permission stations so that control can be regained in the case of compromise or loss~\cite{llansoAchievingSpaceMission2016}. This is one potential use for emerging ``Ground Station as a Service'' offerings such as Amazon Web Services pay-as-you-go Ground Station~\cite{amazonwebservicesAWSGroundStation}.

Second, satellite ground systems may be located in remote areas with limited physical security controls~\cite{bardinSatelliteCyberAttack2013}. This arises because the main placement considerations relate to signal coverage and access to a particular orbit - rather than terrestrial accessibility. Often, little to no staff will have a regular physical presence on-site~\cite{dscSpaceFinalFrontier2016}. Instead, day-to-day operations will be highly automated and controlled remotely from a centralized operations center~\cite{dscSpaceFinalFrontier2016}. This increases the threat of attacks leveraging physical access and contrasts with physical controls applied to many other critical information systems.

Finally, satellite ground stations are generally the main ``bridge'' between the terrestrial internet and satellites. Due to heavy use of remote access, ground stations are difficult to fully ``air-gap''~\cite{laneHighAssuranceCyberSpace2017, cunninghamEffectiveCybersecurityModular2016}. Prior security research has found numerous readily exploitable vulnerabilities in ground-station software and demonstrated that satellite ground terminals can be easily identified using IOT search engines like Shodan~\cite{santamartaLastCallSATCOM2018,santamartaSATCOMTerminalsHacking2014}. Moreover, relative normalcy of ground station hardware means barriers to entry are low compared to other segments.

Generally, traditional enterprise security practices are prescribed to defend ground systems. For example, auditing malware on a satellite ground station can be done with traditional forensic tools~\cite{cohenSpacecraftEmbeddedCyber2016}. There are some systems which are unique to the satellite environment and may require special security treatment - such as long-range radio hardware~\cite{veraCyberSecurityAwareness2016}. However, our historical analysis has found no public instance of attacks targeting this equipment specifically and limited academic study of these components.

In sum, satellite ground station security is typically considered an extension of traditional IT security. The critical difference is often the severity of potential harms rather than mechanisms of attacks and defenses. However, this maxim is far from universal. Future offensive security work focused on unique satellite mission control hardware and software may uncover previously overlooked vulnerabilities.

\section{Holistic Security Models}
While the subsystem taxonomization in this paper is useful for identifying technical challenges and contributions, it neglects one key area of evolution in real-world satellite security practice. In recent years, a sizeable literature base has emerged discussing high-level organizational best practices and security frameworks. As this research tends to be more theoretical than applied, historical barriers to equipment access are acute. This may explain the relative abundance of security frameworks compared to technical research.

Generally, these frameworks can be categorized into two broad classes: those which focus on the organizational practice of satellite operators and those which focus on the duties of policymakers. In this section, we will briefly discuss some of the core challenges facing such frameworks and some of the more consequential proposals.

\subsection{Operational Frameworks}
A key first step in developing any satellite security framework is to define its scope. Cunningham et al. argue that this is best done by dividing satellite missions into five broad phases and linking each phase to a distinct ``cyber-security overlay'' which promotes security by design. For example, in the ``payload and subsystem development'' phase they suggest that satellite operators ``incorporate security code and controls''~\cite{cunninghamEffectiveCybersecurityModular2016}. Like many high-level frameworks, the core technical dimensions here are somewhat vague. However, the phase-oriented approach does bring a key benefit in delineating which organizations are responsible for given protections - a historical challenge discussed at length in Section~\ref{sec:security-challenges}.

An alternative framing is proposed by Zatti in which satellite security controls are tied to specific mission types with the addition of some generic controls common to all missions~\cite{zattiProtectionSpaceMissions2017}. Vivero suggests a similar approach~\cite{viveroSpaceMissionsCybersecurity2013}. This framing seeks to balance the diversity of satellite systems with the need for common best practices. One key advantage of mission-framing is in threat modeling. For example, the attackers interested in harming human spaceflight have radically different capabilities and motivations from those interested in compromising satellite television. Unfortunately, this framing leaves ambiguity in multi-stakeholder projects as to which organization is responsible for implementing which controls.

CCSDS suggests a hybrid approach~\cite{ccsdsSecurityThreatsSpace2015}. This remediates the jurisdictional shortcomings of a pure mission-class approach while providing clearer threat-modeling. The proposal incorporates explicit consideration of mission-based attack probability mapped to lifestyle stages. While this is not presented as exhaustive framework, but rather a proof-of-concept, it is nevertheless among the most technically comprehensive examples to date.

The most commonly suggested approach, however, is to map pre-existing IT security controls to satellite systems, though these suggestions rarely include specific mappings~\cite{knezLessonsLearnedApplying2016,youngCommercialSatellitesCritical2017, veraCyberSecurityAwareness2016}. This is appealing because it draws on a set of generally accepted best practices. However, as noted by Knez et al., the uniqueness of space systems complicates this process and many controls are only superficially meaningful~\cite{knezLessonsLearnedApplying2016}. Such standards neither consider the unique threat models targeting satellite systems nor the multi-stakeholder ecosystem. Moreover, they lack differentiation between the lifecycle stages - assuming relatively static systems. The threats facing a web server are largely consistent throughout its life, but the threats facing a satellite during orbital injection differ radically from those threatening a broadcasting communications platform. In practice, the NIST Cybersecurity Framework is widely employed in industry but it is unclear if it is fit for purpose~\cite{csricCybersecurityRiskManagement2015}.

\subsection{Policy and Legislative Frameworks}

Given the importance of satellites to modern information societies, it has been suggested that satellite operators may not adequately self-regulate. This is especially concerning for dual-use systems which are commercially owned but provide critical communications linkages to government operations. As such, it may be necessary to adopt regulations that re-balance incentive structures to better prioritize security.

One of the primary discussions is taxonomic. The question as to whether or not satellite systems are considered ``critical infrastructure'' remains unsettled and has a significant impact on the way in which companies and governments must protect them~\cite{delmonteCybersecurityPolicySustainable2013}. This may explain the relative paucity of satellite standards compared to similar infrastructure sectors~\cite{falcoVacuumSpaceCyber2018, youngCommercialSatellitesCritical2017}.

A general desire to classify satellites as critical infrastructure has been acknowledged by the US government since at least 2002, however an explicit classification of this nature has yet to occur~\cite{u.s.governmentaccountabilityofficeCriticalInfrastructureProtection2002}. Such classification may force improvements, particularly with regards to redundancy and supply chain verification. However, industry actors have expressed resistance to rigid legal standards, contending that status quo requirements are adequate~\cite{csricCybersecurityRiskManagement2015}.

Beyond the critical infrastructure debate, an additional point of contention regards the legal rights of satellite operators to defend themselves. Rendleman and Ryals suggest satellite operators should be permitted to corrupt files and commit denial of service attacks (e.g. spectrum jamming) against attackers to regain control of their satellites~\cite{rendlemanCyberOperationsDefend2013}. They suggest the use of letters of Marque and Reprisal, a historical practice which allowed privateers to engage in combat against foreign vessels on the high seas~\cite{rendlemanCyberOperationsDefend2013}. This aligns with a broader trend applying maritime policy frameworks to space~\cite{meyerOuterSpaceCyber2016}. However, such ``hack-back'' rights are highly controversial~\cite{rendlemanCyberOperationsDefend2013}.

One final notable genre of policy development centers on the international dynamics of satellite cyber-security. Housen-Couriel argues that status quo practice has created a legal lacuna in which it is unclear which international organizations and laws apply to satellite hacking incidents~\cite{housen-courielCybersecurityAntiSatelliteCapabilities2014}. This suggests a need for new international law that either clarifies the applicability of existing frameworks or the creates new frameworks specific to space systems~\cite{housen-courielCybersecurityAntiSatelliteCapabilities2014}. Chatham House makes similar suggestions, pointing to the International Telecommunications Union (ITU) as the ideal regulatory body for such a regime~\cite{dscSpaceFinalFrontier2016}. They further suggest that this should incorporate interstate threat intelligence sharing due to trans-national effects of satellite failure - something which has been historically constrained by high classification levels~\cite{dscSpaceFinalFrontier2016}. Blount contends that cyber-intelligence sharing in space has promise due to existing collaborations (such as on debris tracking)~\cite{blountSatellitesAreJust2017}. While little progress has been made thus far, it remains possible that policymakers will seek technical input into the design of such systems.

Ultimately, we find a substantial body of policy research which has evolved more or less in isolation from relevant technical communities. The result is that many proposals appear aspirational rather than actionable. Much as in other areas, conscious effort by the system security community to bridge this gap may pave the way for novel and impactful future work in both fields.

\section{Conclusion}
Satellites are an increasingly vital component of modern life and their security represents a key point of failure in systems ranging from military communications to meteorological forecasting. Our analysis of 60 years of historical trends suggest that satellites will continue facing sophisticated, aggressive, and constantly evolving threats in cyberspace.

Despite this legacy, the intersection between outer space and cyberspace remains poorly understood - with important contributions scattered across diverse and isolated disciplines. In this paper, we have synthesized these perspectives to draw out research problems which the systems security community can contribute towards solving.

In the communications domain, we find a need for substantial cryptographic developments to provide secure and commercially palatable alternatives to dominant satellite radio protocols. With respect to satellite platforms, we find that almost no technical research exists on the defense and monitoring of systems in orbit - especially against ground-inserted malware. On the ground, we find that general IT security approaches are popular among satellite operators, but that little research has consider unique functions of space control software or signaling hardware. Finally, from a high-level operational perspective, we find many aspirational mission security framework proposals, but little research which maps policy objectives to clear technical practices and implementations.

As thousands of satellites reach orbit over the next decade, these questions cannot remain unanswered. There is a critical opportunity for the systems security community to build upon the research of others and collaborate to better protect the next half-century of human spaceflight.

\printbibliography
\appendices	
%\begin{landscape}
\onecolumn
\section{Satellite Security Incident Chronology}
\begin{longtabu}{l X X X X X}
	Year & Attack \newline Type & Attacker \newline Type & Attacker \newline Country & Victim \newline Type & Victim \newline Country\\
	\hline
	\hline
	\\
	\endhead
	1962 & Jamming & Government & Hypothetical & Commercial & United States\\*
	& \multicolumn{5}{p{\dimexpr 5\tabucolX+5\tabcolsep+\arrayrulewidth\relax}}{In a 1962 congressional hearing on the first American commercial satellite company, the prospect of signal jamming and potential satellite hijacking was suggested as a possible threat to low-altitude satellite missions.}\\&\multicolumn{5}{p{\dimexpr 5\tabucolX+5\tabcolsep+\arrayrulewidth\relax}}{\textit{Primary/Contemporary References}:~\cite{newyorktimesRadioSatellitesOpen1962}}\\\\
	
	1972 & Jamming & Government & Soviet Union & Multiple & Multiple\\*
	& \multicolumn{5}{p{\dimexpr 5\tabucolX+5\tabcolsep+\arrayrulewidth\relax}}{A UN proposal by the Soviet Union is raised suggesting that states have an intrinsic right to jam satellite signals in their territories via technical means.}\\&\multicolumn{5}{p{\dimexpr 5\tabucolX+5\tabcolsep+\arrayrulewidth\relax}}{\textit{Primary/Contemporary References}:~\cite{washingtonpostSovietsAskControls1972} }\\\\
	
	1986 & Signal Hijacking & Insider & United States & Commercial & United States\\*
	& \multicolumn{5}{p{\dimexpr 5\tabucolX+5\tabcolsep+\arrayrulewidth\relax}}{An industry insider injected video and audio into an HBO television broadcast in Florida. Interestingly, this attack may have been inspired by a fictional article which appeared the previous year in a satellite television enthusiast magazine about an individual who hijacked HBO signals in protest of new scrambling policies (Pollack 1986).}\\&\multicolumn{5}{p{\dimexpr 5\tabucolX+5\tabcolsep+\arrayrulewidth\relax}}{\textit{Primary/Contemporary References}:~\cite{shalesCableCaptainMidnight1986,davisCaptainMidnightUnmasked1986}~\textit{Secondary References}:~\cite{youngCommercialSatellitesCritical2017,fritzSatelliteHackingGuide2013}}\\\\
	
	1986 & Jamming & Government & United States & Ground  (Accidental) & United States\\*
	& \multicolumn{5}{p{\dimexpr 5\tabucolX+5\tabcolsep+\arrayrulewidth\relax}}{In 1986 a garage door company discovered that communications satellites which were directed towards Regan's vacation home in California were jamming terrestrial garage door openers more than 200 miles away.}\\&\multicolumn{5}{p{\dimexpr 5\tabucolX+5\tabcolsep+\arrayrulewidth\relax}}{\textit{Primary/Contemporary References}:~\cite{washingtonpostDoorsFailWhen1986}}\\\\
	
	1986 & Eavesdropping & Government & Indonesia & Commercial & United States\\*
	& \multicolumn{5}{p{\dimexpr 5\tabucolX+5\tabcolsep+\arrayrulewidth\relax}}{In 1986 the government of Indonesia was accused by an American satellite imaging firm of using large satellite receivers to intercept earth observation images without subscribing to the service.}\\&\multicolumn{5}{p{\dimexpr 5\tabucolX+5\tabcolsep+\arrayrulewidth\relax}}{\textit{Primary/Contemporary References}:~\cite{timesofindiaSpacePiracy1986}}\\\\
	
	1987 & Signal Hijacking & Individual & United States & Commercial & United States\\*
	& \multicolumn{5}{p{\dimexpr 5\tabucolX+5\tabcolsep+\arrayrulewidth\relax}}{In 1987, Thomas Haynie, an employee of the Christian Broadcasting Network hijacked satellite transmissions from the Playboy Channel and replaced them with static text from the bible.}\\&\multicolumn{5}{p{\dimexpr 5\tabucolX+5\tabcolsep+\arrayrulewidth\relax}}{\textit{Primary/Contemporary References}:~\cite{smithAppealUnitedStates1991}~\textit{Secondary References}:~\cite{fritzSatelliteHackingGuide2013}}\\\\
	
	1987 & Groundstation & Individual & Germany & Gov. Military & United States\\*
	& \multicolumn{5}{p{\dimexpr 5\tabucolX+5\tabcolsep+\arrayrulewidth\relax}}{In 1987, a group of youths in West Germany managed to compromise top secret networks belonging to NASA and other major space agencies. These networks provided at least the ability to find secret information about space missions and potentially information which could have compromised these missions.}\\&\multicolumn{5}{p{\dimexpr 5\tabucolX+5\tabcolsep+\arrayrulewidth\relax}}{\textit{Primary/Contemporary References}:~\cite{parryYouthsHackedSecret1987}}\\\\
	
	1993 & Cryptographic & Individual & United Kingdom & Commercial & United Kingdom\\*
	& \multicolumn{5}{p{\dimexpr 5\tabucolX+5\tabcolsep+\arrayrulewidth\relax}}{A group of hackers distributed BSkyB satellite channels through a decoder-card sharing scheme across an apartment complex.}\\&\multicolumn{5}{p{\dimexpr 5\tabucolX+5\tabcolsep+\arrayrulewidth\relax}}{\textit{Primary/Contemporary References}:~\cite{hellenHackersTargetBskyB1993}}\\\\
	
	1993 & Jamming & Government & Indonesia & Commercial & Tonga\\*
	& \multicolumn{5}{p{\dimexpr 5\tabucolX+5\tabcolsep+\arrayrulewidth\relax}}{Satellite operators from Indonesia and Tongo threatened each other over the proposed Tonga Gorizont 17 satellite in GEO above New Guinea. The orbital slot was under contention due to potential interference as both states threatened to jam the other's transmissions from the orbit. This is the first public record of a state threatening digital counterspace operations against another state's assets.}\\&\multicolumn{5}{p{\dimexpr 5\tabucolX+5\tabcolsep+\arrayrulewidth\relax}}{\textit{Primary/Contemporary References}:~\cite{steinElbowingPieceSpace1993}}\\\\
	
	1994 & Jamming & Government & Iran & Commercial & Multiple\\*
	& \multicolumn{5}{p{\dimexpr 5\tabucolX+5\tabcolsep+\arrayrulewidth\relax}}{The Iranian government was suspected of jamming foreign television programs from Arab-Sat and Asia-Sat platforms during ongoing debate over banning the domestic use of satellite dishes altogether.}\\&\multicolumn{5}{p{\dimexpr 5\tabucolX+5\tabcolsep+\arrayrulewidth\relax}}{\textit{Primary/Contemporary References}:~\cite{afpForeignSatellitePrograms1994}}\\\\
	
	1994 & Cryptographic & Individual & United States & Commercial & United States\\*
	& \multicolumn{5}{p{\dimexpr 5\tabucolX+5\tabcolsep+\arrayrulewidth\relax}}{Gregory Manzer was sentenced on charges of creating and distributing technology to break the VideoCipher encryption technology used by HBO and ESPN satellite channels.}\\&\multicolumn{5}{p{\dimexpr 5\tabucolX+5\tabcolsep+\arrayrulewidth\relax}}{\textit{Primary/Contemporary References}:~\cite{walshCableTVPirate1994}}\\\\
	
	1996 & Jamming & Government & Indonesia & Commercial & Hong Kong\\*
	& \multicolumn{5}{p{\dimexpr 5\tabucolX+5\tabcolsep+\arrayrulewidth\relax}}{In 1996 the Indonesian government used a communications satellite called Palapa B1 to jam signals from a Hong Kong/British satellite Apstar-1A which was leased by Tonga - making good on threats from three years prior.}\\&\multicolumn{5}{p{\dimexpr 5\tabucolX+5\tabcolsep+\arrayrulewidth\relax}}{\textit{Primary/Contemporary References}: N/A~\textit{Secondary References}:~\cite{wongMilitarySpacePower2010,fritzSatelliteHackingGuide2013,manulisCyberSecurityNew2020}}\\\\
	
	1996 & Jamming & Government & Turkey & Commercial & United States\\*
	& \multicolumn{5}{p{\dimexpr 5\tabucolX+5\tabcolsep+\arrayrulewidth\relax}}{The Turkish Government is believed to have jammed broadcasts originating from MED-TV, a Kurdish nationalist satellite television station operating on an Eutelsat satellite. The jamming campaign continued sporadically between 1996 and 1999. Turkish authorities claimed MED-TV was ``Terrorist Television'' and incited acts of violence. British authorities ultimately terminated the MED-TV transponder license in 1999.}\\&\multicolumn{5}{p{\dimexpr 5\tabucolX+5\tabcolsep+\arrayrulewidth\relax}}{\textit{Primary/Contemporary References}:~\cite{boustanyKurdishTVGets1998,kinzerKurdsAreDetermined1999}~\textit{Secondary References}:~\cite{hassanpourSatelliteFootprintsNational1998,wongMilitarySpacePower2010}}\\\\
	
	1997 & Groundstation & Government & Russia & Gov. Scientific & United States \\*
	& \multicolumn{5}{p{\dimexpr 5\tabucolX+5\tabcolsep+\arrayrulewidth\relax}}{Hackers in 1997 successfully compromised Goddard Space Flight Center computers capable of satellite command and control. Later investigation linked this incident to Russia-government associated hackers although full verification of this claim cannot be made without access to classified investigation reports.}\\&\multicolumn{5}{p{\dimexpr 5\tabucolX+5\tabcolsep+\arrayrulewidth\relax}}{\textit{Primary/Contemporary References}:~\cite{elginNetworkSecurityBreaches2008}~\textit{Secondary References}:~\cite{fritzSatelliteHackingGuide2013}}\\\\
	
	1998 & Payload Damage & Unknown & Unknown & Gov. Scientific & United States\\*
	& \multicolumn{5}{p{\dimexpr 5\tabucolX+5\tabcolsep+\arrayrulewidth\relax}}{A cyber-intrusion at Goddard Space Flight Center possibly caused the German-US ROSAT telescope to face the sun and burn its optical sensors.}\\&\multicolumn{5}{p{\dimexpr 5\tabucolX+5\tabcolsep+\arrayrulewidth\relax}}{\textit{Primary/Contemporary References}:~\cite{elginNetworkSecurityBreaches2008}~\textit{Secondary References}:~\cite{zattiProtectionSpaceMissions2017}}\\\\
	
	1998 & Jamming & Commercial & Russia & Multiple & Multiple\\*
	& \multicolumn{5}{p{\dimexpr 5\tabucolX+5\tabcolsep+\arrayrulewidth\relax}}{A Moscow-based company began selling a {\$}4000 portable jammer capable of disabling GPS signals over a 200km radius.}\\&\multicolumn{5}{p{\dimexpr 5\tabucolX+5\tabcolsep+\arrayrulewidth\relax}}{\textit{Primary/Contemporary References}:~\cite{intelligencenewsletterAnybodyNeedGPS1998}}\\\\
	
	1998 & Groundstation & Individual & United States & Gov. Military & United States\\*
	& \multicolumn{5}{p{\dimexpr 5\tabucolX+5\tabcolsep+\arrayrulewidth\relax}}{In 1998, a hacker group called ''Masters of Downloading'' claimed to have stolen classified software that provided sensitive information and limited control over military satellites include GPS systems. The pentagon acknowledged a minor breach but contended that the hackers exaggerated their capabilities.}\\&\multicolumn{5}{p{\dimexpr 5\tabucolX+5\tabcolsep+\arrayrulewidth\relax}}{\textit{Primary/Contemporary References}:~\cite{glaveHaveCrackersFound1998}~\textit{Secondary References}:~\cite{fritzSatelliteHackingGuide2013}}\\\\
	
	1999 & TT{\&}C & Individual & Unknown & Gov. Military & United Kingdom\\*
	& \multicolumn{5}{p{\dimexpr 5\tabucolX+5\tabcolsep+\arrayrulewidth\relax}}{In 1999 hackers claimed to have hijacked a British military satellite's control systems and to have demanded ransom from the British government. However, the British military strongly disputed these claims.}\\&\multicolumn{5}{p{\dimexpr 5\tabucolX+5\tabcolsep+\arrayrulewidth\relax}}{\textit{Primary/Contemporary References}:~\cite{bbcSatelliteHijackImpossible1999}~\textit{Secondary References}:~\cite{fritzSatelliteHackingGuide2013}}\\\\
	
	1999 & Jamming & Government & Russia & Commercial & Russia\\*
	& \multicolumn{5}{p{\dimexpr 5\tabucolX+5\tabcolsep+\arrayrulewidth\relax}}{Russian government admits jamming satellite phone networks in Chechnya to present communications among separatists.}\\&\multicolumn{5}{p{\dimexpr 5\tabucolX+5\tabcolsep+\arrayrulewidth\relax}}{\textit{Primary/Contemporary References}:~\cite{afpMoscowAdmitsSatellite1999}}\\\\
	
	1999 & Groundstation & Individual & United States & Gov. Manned & Multiple\\*
	& \multicolumn{5}{p{\dimexpr 5\tabucolX+5\tabcolsep+\arrayrulewidth\relax}}{A teenager going by the name cOmrade plead guilty to charges of compromising NASA computer systems that support the International Space Station. The intrusions occurred in 1999.}\\&\multicolumn{5}{p{\dimexpr 5\tabucolX+5\tabcolsep+\arrayrulewidth\relax}}{\textit{Primary/Contemporary References}:~\cite{wilsonTeenGivenSix2000}}\\\\
	
	2000 & Jamming & Government & France & Navigational & Multiple\\*
	& \multicolumn{5}{p{\dimexpr 5\tabucolX+5\tabcolsep+\arrayrulewidth\relax}}{During a 2000 tank competition to demonstrate tanks for sale to the Greek military, French forces used ground-based GPS jammers to cause navigation problems during US and British entries.}\\&\multicolumn{5}{p{\dimexpr 5\tabucolX+5\tabcolsep+\arrayrulewidth\relax}}{ ~\textit{Secondary References}:~\cite{grauGPSSignalsJammed2001,fritzSatelliteHackingGuide2013}}\\\\
	
	2000 & Jamming & Government & Iran & Commercial & France\\*
	& \multicolumn{5}{p{\dimexpr 5\tabucolX+5\tabcolsep+\arrayrulewidth\relax}}{Iran accused of using jamming devices impacting Turkish territory to interfere with Eutelsat-based opposition broadcasts.}\\&\multicolumn{5}{p{\dimexpr 5\tabucolX+5\tabcolsep+\arrayrulewidth\relax}}{\textit{Primary/Contemporary References}:~\cite{afpIranianGovernmentJamming2000}}\\\\
	
	2000 & Groundstation & Individual & United States & Gov. Scientific & United States\\*
	& \multicolumn{5}{p{\dimexpr 5\tabucolX+5\tabcolsep+\arrayrulewidth\relax}}{Jason Dikeman was charged in 2000 of gaining unauthorized access to systems which control NASA satellites.}\\&\multicolumn{5}{p{\dimexpr 5\tabucolX+5\tabcolsep+\arrayrulewidth\relax}}{\textit{Primary/Contemporary References}:~\cite{costelloSuspectArrestedNASA2000}}\\\\
	
	2000 & Groundstation & Unknown & Unknown & Gov. Military & United States\\*
	& \multicolumn{5}{p{\dimexpr 5\tabucolX+5\tabcolsep+\arrayrulewidth\relax}}{In 2000 unknown hackers stole software from a US defense contracting company which enables ground stations to send commands to satellites.}\\&\multicolumn{5}{p{\dimexpr 5\tabucolX+5\tabcolsep+\arrayrulewidth\relax}}{\textit{Primary/Contemporary References}:~\cite{lemosSatelliteControlCodes2001}}\\\\
	
	2001 & Groundstation & Individual & United Kingdom & Gov. Scientific & United States\\*
	& \multicolumn{5}{p{\dimexpr 5\tabucolX+5\tabcolsep+\arrayrulewidth\relax}}{UK based hacker Gary McKinnon was indicted on charges of compromising 16 NASA computer systems. McKinnon claimed to have been looking for evidence of a cover-up relating to extra-terrestrial intelligence and unidentified flying objects. While there is not evidence that McKinnon compromised systems related to satellite control, it represents an early high-profile attack against a space agency with the intent of stealing space-mission data. Subsequent coverage has focused on matters of extradition and human rights for cyber-crime.}\\&\multicolumn{5}{p{\dimexpr 5\tabucolX+5\tabcolsep+\arrayrulewidth\relax}}{\textit{Primary/Contemporary References}:~\cite{mcnultyUnitedStatesAmerica2002}~\textit{Secondary References}:~\cite{mckinnonTheresaMaySaved2016,manulisCyberSecurityNew2020}}\\\\
	
	2002 & Groundstation & Individual & Venezuela & Gov. Scientific & United States\\*
	& \multicolumn{5}{p{\dimexpr 5\tabucolX+5\tabcolsep+\arrayrulewidth\relax}}{A Venezuelan hacker using the pseudonym ''RaFa'' provided a reporter at Computer World copies of a PowerPoint documents detailing the design of NASA launch vehicle Cobra and other sensitive engineering information. Later, Rafael N{\~A}{\textordmasculine}{\~A}{\ensuremath{\pm}}ez Aponte was sentenced and extradited for compromises and defacement of US military information systems conducted under the same pseudonym but charges for the NASA compromise were never pressed. Some sources associate this compromise with the 2002 Marshal Space Flight Center Intrusions, although this attribution is disputed.}\\&\multicolumn{5}{p{\dimexpr 5\tabucolX+5\tabcolsep+\arrayrulewidth\relax}}{\textit{Primary/Contemporary References}:~\cite{schwartzCompressedDataHacker2002}~\textit{Secondary References}:~\cite{elginNetworkSecurityBreaches2008}}\\\\
	
	2002 & Signal Hijacking & Dissident & Taiwan & Gov. Media & China\\*
	& \multicolumn{5}{p{\dimexpr 5\tabucolX+5\tabcolsep+\arrayrulewidth\relax}}{Falun Gong transmitted protest videos from Taipei over official Chinese Central Television satellite broadcasts.}\\&\multicolumn{5}{p{\dimexpr 5\tabucolX+5\tabcolsep+\arrayrulewidth\relax}}{\textit{Primary/Contemporary References}:~\cite{associatedpressFalunGongHijacks2002}}\\\\
	
	2002 & Groundstation & Government & China & Gov. Scientific & United States\\*
	& \multicolumn{5}{p{\dimexpr 5\tabucolX+5\tabcolsep+\arrayrulewidth\relax}}{An attacker compromised computers at Marshall Space Flight Center stealing intellectual property related to launch vehicle design. This attack has since been tenuously attributed to China, although more contemporaneous sources associate it with the Rafa intrusions.}\\&\multicolumn{5}{p{\dimexpr 5\tabucolX+5\tabcolsep+\arrayrulewidth\relax}}{\textit{Primary/Contemporary References}:~\cite{schwartzCompressedDataHacker2002} ~\textit{Secondary References}:~\cite{elginNetworkSecurityBreaches2008,fritzSatelliteHackingGuide2013}}\\\\
	
	2002 & Jamming & Individual & United Kingdom & Navigational & United States\\*
	& \multicolumn{5}{p{\dimexpr 5\tabucolX+5\tabcolsep+\arrayrulewidth\relax}}{Several sources assert that in 2002, a poorly installed CCTV camera in the town of Douglas, Isle of Mann caused interference with GPS signals over a 1 km area. We were unable to find a primary source for this claim, but it is a commonly referenced example of accident GPS interference.}\\&\multicolumn{5}{p{\dimexpr 5\tabucolX+5\tabcolsep+\arrayrulewidth\relax}}{ ~\textit{Secondary References}:~\cite{martinsGNSSVulnerabilitesRobustness2014,royalacademyofengineeringGlobalNaviationSpace2011,manulisCyberSecurityNew2020}}\\\\
	
	2002 & Eavesdropping & Individual & United Kingdom & Gov. Military & United States\\*
	& \multicolumn{5}{p{\dimexpr 5\tabucolX+5\tabcolsep+\arrayrulewidth\relax}}{John Locker, a satellite eavesdropper, reported the ability to intercept images from NATO surveillance aircraft. NATO respondents claimed that the images did not contain sensitive information but media reports claimed that they revealed sensitive details regarding the capabilities and location of classified vehicles.}\\&\multicolumn{5}{p{\dimexpr 5\tabucolX+5\tabcolsep+\arrayrulewidth\relax}}{\textit{Primary/Contemporary References}:~\cite{urbanEnthusiastWatchesNato2002}~\textit{Secondary References}:~\cite{manulisCyberSecurityNew2020}}\\\\
	
	2003 & Signal Jamming & Government & Iran & Commercial & United States\\*
	& \multicolumn{5}{p{\dimexpr 5\tabucolX+5\tabcolsep+\arrayrulewidth\relax}}{US government broadcasts in favor of regime change in Iran were jammed by attacks on the Telstar-12 satellites by the Iranian government.}\\&\multicolumn{5}{p{\dimexpr 5\tabucolX+5\tabcolsep+\arrayrulewidth\relax}}{\textit{Primary/Contemporary References}:~\cite{windremSatelliteFeedsIran2003}~\textit{Secondary References}:~\cite{littmanSatelliteNetworkSecurity2009}}\\\\
	
	2003 & Jamming & Government & Cuba & Commercial & United States\\*
	& \multicolumn{5}{p{\dimexpr 5\tabucolX+5\tabcolsep+\arrayrulewidth\relax}}{In 2003 the Cuban government was accused of deliberately jamming US signals for the Voice of America station which were being broadcast to Iran, perhaps on behalf of the Iranian government and with communications gear supplied by China. This has been associated with the Iranian jamming of the Telstar-12 incident by some secondary sources.}\\&\multicolumn{5}{p{\dimexpr 5\tabucolX+5\tabcolsep+\arrayrulewidth\relax}}{\textit{Primary/Contemporary References}:~\cite{marquezUSCondemnsCuba2003}~\textit{Secondary References}:~\cite{manulisCyberSecurityNew2020}}\\\\
	
	2003 & Signal Hijacking & Dissident & Taiwan & Commercial & China\\*
	& \multicolumn{5}{p{\dimexpr 5\tabucolX+5\tabcolsep+\arrayrulewidth\relax}}{Falun Gong again transmitted protest media across an AsiaSat transponder in 2003 to interrupt CCTV coverage of the Zhenzhou V space mission.}\\&\multicolumn{5}{p{\dimexpr 5\tabucolX+5\tabcolsep+\arrayrulewidth\relax}}{\textit{Primary/Contemporary References}:~\cite{southchinamorningpostFalunGongAccused2003}}\\\\
	
	2004 & Signal Hijacking & Dissident & Taiwan & Commercial & China\\*
	& \multicolumn{5}{p{\dimexpr 5\tabucolX+5\tabcolsep+\arrayrulewidth\relax}}{Falun Gong again transmitted protest media across an AsiaSat transponder.}\\&\multicolumn{5}{p{\dimexpr 5\tabucolX+5\tabcolsep+\arrayrulewidth\relax}}{\textit{Primary/Contemporary References}:~\cite{xinhuaAsiaSatAccusesFalungong2004}}\\\\
	
	2005 & Groundstation & Government & China & Gov. Manned & United States\\*
	& \multicolumn{5}{p{\dimexpr 5\tabucolX+5\tabcolsep+\arrayrulewidth\relax}}{Windows malware installed in Kennedy space center's vehicle assembly building sent information about the space shuttle to computers in Taiwan. While this may have been espionage to mimic shuttle technology, investigators also believe information that could threaten the shuttle was exfiltrated. Weak attribution to the PLA has been made.}\\&\multicolumn{5}{p{\dimexpr 5\tabucolX+5\tabcolsep+\arrayrulewidth\relax}}{ ~\textit{Secondary References}:~\cite{elginNetworkSecurityBreaches2008,fritzSatelliteHackingGuide2013}}\\\\
	
	2005 & Jamming & Government & Libya & Commercial & United States\\*
	& \multicolumn{5}{p{\dimexpr 5\tabucolX+5\tabcolsep+\arrayrulewidth\relax}}{In 2005 the Libyan government was accused of jamming telecommunications satellites which impacted both European television stations and government communications.}\\&\multicolumn{5}{p{\dimexpr 5\tabucolX+5\tabcolsep+\arrayrulewidth\relax}}{\textit{Primary/Contemporary References}:~\cite{henckeProtestLibyaSatellites2005}~\textit{Secondary References}:~\cite{fritzSatelliteHackingGuide2013}}\\\\
	
	2006 & Jamming & Government & Libya & Commercial & United Arab Emirates\\*
	& \multicolumn{5}{p{\dimexpr 5\tabucolX+5\tabcolsep+\arrayrulewidth\relax}}{In 2006 the Libyan government was accused of jamming satellite telephone frequencies in order to combat the use of satphones by smugglers.}\\&\multicolumn{5}{p{\dimexpr 5\tabucolX+5\tabcolsep+\arrayrulewidth\relax}}{\textit{Primary/Contemporary References}:~\cite{choiLibyaPinpointedSource2007}~\textit{Secondary References}:~\cite{fritzSatelliteHackingGuide2013}}\\\\
	
	2006 & Signal Hijacking & Government & Israel & Commercial & Lebanon\\*
	& \multicolumn{5}{p{\dimexpr 5\tabucolX+5\tabcolsep+\arrayrulewidth\relax}}{Israeli forces in 2006 hijacked the Hezbollah-associated satellite television channels to air threatening anti-Hezbollah messages.}\\&\multicolumn{5}{p{\dimexpr 5\tabucolX+5\tabcolsep+\arrayrulewidth\relax}}{\textit{Primary/Contemporary References}:~\cite{spectorHackingHezbollah2006}~\textit{Secondary References}:~\cite{fritzSatelliteHackingGuide2013}}\\\\
	
	2006 & Groundstation & Unknown & Unknown & Gov. Scientific & United States\\*
	& \multicolumn{5}{p{\dimexpr 5\tabucolX+5\tabcolsep+\arrayrulewidth\relax}}{A purported 2006 phishing incident targeting NASA employees lead to the leak of NASA budgetary documents detailing satellite investment priorities. We were unable to find primary source information regarding this breach, but several prior surveys have cited it as example of IP theft attacks.}\\&\multicolumn{5}{p{\dimexpr 5\tabucolX+5\tabcolsep+\arrayrulewidth\relax}}{~\textit{Secondary References}:~\cite{elginNetworkSecurityBreaches2008,fritzSatelliteHackingGuide2013,manulisCyberSecurityNew2020}}\\\\
	
	2006 & Sensor Disruption & Government & China & Gov. Military & United States\\*
	& \multicolumn{5}{p{\dimexpr 5\tabucolX+5\tabcolsep+\arrayrulewidth\relax}}{China beamed a ground-based laser at sensors on a US spy satellite. Very little information about the incident and its effects is public.}\\&\multicolumn{5}{p{\dimexpr 5\tabucolX+5\tabcolsep+\arrayrulewidth\relax}}{\textit{Primary/Contemporary References}:~\cite{couriermailChinaTargetsUS2006}}\\\\
	
	2007 & Signal Hijacking & Terrorist & Sri Lanka & Commercial & United States\\*
	& \multicolumn{5}{p{\dimexpr 5\tabucolX+5\tabcolsep+\arrayrulewidth\relax}}{Tamil rebels may have hijacked an Intelsat satellite signal to broadcast propaganda. The rebels claim they had purchased access to the satellite but Intelsat disputes this. The incident went on for more than 2 years.}\\&\multicolumn{5}{p{\dimexpr 5\tabucolX+5\tabcolsep+\arrayrulewidth\relax}}{\textit{Primary/Contemporary References}:~\cite{dalyLTTETechnologicallyInnovative2007}~\textit{Secondary References}:~\cite{bardinSatelliteCyberAttack2013}}\\\\
	
	2007 & Groundstation & Government & China & Gov. Scientific & United States\\*
	& \multicolumn{5}{p{\dimexpr 5\tabucolX+5\tabcolsep+\arrayrulewidth\relax}}{The ground station analysis process for Earth Observation Data at Goddard Space Flight center was compromised by attackers believed to be associated with the Chinese state according to secondary sources. No primary source coverage of this incident could be found, but it is cited in several surveys as an instance of state sponsored espionage.}\\&\multicolumn{5}{p{\dimexpr 5\tabucolX+5\tabcolsep+\arrayrulewidth\relax}}{\textit{Primary/Contemporary References}:~\textit{Secondary References}:~\cite{elginNetworkSecurityBreaches2008,fritzSatelliteHackingGuide2013,manulisCyberSecurityNew2020}}\\\\
	
	2007 & TT{\&}C & Government & China & Gov. Scientific & United States\\*
	& \multicolumn{5}{p{\dimexpr 5\tabucolX+5\tabcolsep+\arrayrulewidth\relax}}{Two NASA satellites in 2007 and 2008 suffered major disruption attacks. Initial reporting suggested that these were just jamming attacks but later reports suggest ground station control takeover and accuse China.}\\&\multicolumn{5}{p{\dimexpr 5\tabucolX+5\tabcolsep+\arrayrulewidth\relax}}{\textit{Primary/Contemporary References}:~\cite{arthurChineseHackersSuspected2011}~\textit{Secondary References}:~\cite{bardinSatelliteCyberAttack2013,zattiProtectionSpaceMissions2017}}\\\\
	
	2008 & Payload Damage & Insider & Russia & Gov. Manned & Multiple\\*
	& \multicolumn{5}{p{\dimexpr 5\tabucolX+5\tabcolsep+\arrayrulewidth\relax}}{An astronaut is believed to have introduced a virus to ISS windows-XP computers by bringing a compromised laptop on board. More recent reports suggest the virus was brought aboard by Russian cosmonauts, but it is unlikely to have been done deliberately.}\\&\multicolumn{5}{p{\dimexpr 5\tabucolX+5\tabcolsep+\arrayrulewidth\relax}}{\textit{Primary/Contemporary References}:~\cite{francisComputerVirusInfects2008}~\textit{Secondary References}:~\cite{gibbsInternationalSpaceStation2013,zattiProtectionSpaceMissions2017}}\\\\
	
	2008 & Groundstation & Unknown & Unknown & Gov. Manned & United States\\*
	& \multicolumn{5}{p{\dimexpr 5\tabucolX+5\tabcolsep+\arrayrulewidth\relax}}{Attackers were reported as having used a Trojan horse installed on devices at NASA's Johnson Space Center to compromise communications to the international space station and disrupt some services on-board. It is unclear if the attack was targeted or coincidental.}\\&\multicolumn{5}{p{\dimexpr 5\tabucolX+5\tabcolsep+\arrayrulewidth\relax}}{~\textit{Secondary References}:~\cite{steinbergerSurveySatelliteCommunications2008,fritzSatelliteHackingGuide2013,manulisCyberSecurityNew2020}}\\\\
	
	2009 & Groundstation & Individual & Italy & Gov. Scientific & United States\\*
	& \multicolumn{5}{p{\dimexpr 5\tabucolX+5\tabcolsep+\arrayrulewidth\relax}}{In March 2009, an Italian hacker compromised several NASA systems including systems used to control NASA's Deep Space Network and control systems in Goddard Space Flight Center. NASA claims that no critical harm was posed to space missions.}\\&\multicolumn{5}{p{\dimexpr 5\tabucolX+5\tabcolsep+\arrayrulewidth\relax}}{\textit{Primary/Contemporary References}:~\cite{martinNASACybersecurityExamination2012}}\\\\
	
	2009 & Eavesdropping & Terrorist & Iraq & Gov. Military & United States\\*
	& \multicolumn{5}{p{\dimexpr 5\tabucolX+5\tabcolsep+\arrayrulewidth\relax}}{Iraqi insurgents intercepted unencrypted video streams via satellite links using a commercial software product called SkyGrabber.}\\&\multicolumn{5}{p{\dimexpr 5\tabucolX+5\tabcolsep+\arrayrulewidth\relax}}{\textit{Primary/Contemporary References}:~\cite{gormanInsurgentsHackDrones2009}~\textit{Secondary References}:~\cite{bardinSatelliteCyberAttack2013}}\\\\
	
	2009 & Eavesdropping & Researcher & United Kingdom & Commercial & United Kingdom\\*
	& \multicolumn{5}{p{\dimexpr 5\tabucolX+5\tabcolsep+\arrayrulewidth\relax}}{A 2009 Blackhat presentation demonstrates the ability to intercept live video feeds from DVB-S signals, including sensitive military and media feeds by modifying existing satellite hardware.}\\&\multicolumn{5}{p{\dimexpr 5\tabucolX+5\tabcolsep+\arrayrulewidth\relax}}{\textit{Primary/Contemporary References}:~\cite{laurieAtelliteHackingFun2009}}\\\\
	
	2009 & Signal Hijacking & Individual & Brazil & Gov. Military & United States\\*
	& \multicolumn{5}{p{\dimexpr 5\tabucolX+5\tabcolsep+\arrayrulewidth\relax}}{In 2009 almost 40 individuals in Brazil were arrested on charges of hijacking UHF frequencies belonging to US Naval satellites for personal usage. UHF transponder hijacking is believed to be widely used by criminal organizations and individuals seeking free long-range communications services in remote parts of the country.}\\&\multicolumn{5}{p{\dimexpr 5\tabucolX+5\tabcolsep+\arrayrulewidth\relax}}{\textit{Primary/Contemporary References}:~\cite{wiredGreatBrazilianSatHack2009}~\textit{Secondary References}:~\cite{fritzSatelliteHackingGuide2013}}\\\\
	
	2009 & Jamming & Government & Egypt & Commercial & United Kingdom\\*
	& \multicolumn{5}{p{\dimexpr 5\tabucolX+5\tabcolsep+\arrayrulewidth\relax}}{In May of 2009 the Al-Hiwar satellite station broadcast from the United Kingdom was jammed. No culprit has been conclusively identified but the Egyptian government is strongly suspected.}\\&\multicolumn{5}{p{\dimexpr 5\tabucolX+5\tabcolsep+\arrayrulewidth\relax}}{\textit{Primary/Contemporary References}:~\cite{bbcmonitoringworldmediaUKbasedAlHiwarSatellite2009}}\\\\
	
	2010 & Groundstation & Individual & China & Gov. Scientific & United States\\*
	& \multicolumn{5}{p{\dimexpr 5\tabucolX+5\tabcolsep+\arrayrulewidth\relax}}{A Chinese hacker was arrested on charges of stealing export-controlled data from NASA computer systems. The hacker was arrested by Chinese authorities with supporting evidence provided by the United States. It represents one of the first cooperative law enforcement actions regarding government systems compromise between the two states.}\\&\multicolumn{5}{p{\dimexpr 5\tabucolX+5\tabcolsep+\arrayrulewidth\relax}}{\textit{Primary/Contemporary References}:~\cite{martinNASACybersecurityExamination2012}}\\\\
	
	2010 & Eavesdropping & Researcher & Spain & Commercial & Spain\\*
	& \multicolumn{5}{p{\dimexpr 5\tabucolX+5\tabcolsep+\arrayrulewidth\relax}}{A 2010 Blackhat presentation demonstrates the ability to intercept live internet feeds from DVB-S signals using general purpose equipment}\\&\multicolumn{5}{p{\dimexpr 5\tabucolX+5\tabcolsep+\arrayrulewidth\relax}}{\textit{Primary/Contemporary References}:~\cite{egeaPlayingSatelliteEnvironment2010}}\\\\
	
	2010 & Groundstation & Accidental & United States & Navigational & United States\\*
	& \multicolumn{5}{p{\dimexpr 5\tabucolX+5\tabcolsep+\arrayrulewidth\relax}}{In 2010 an Air Force update to GPS ground control stations resulted in multi-day outages effecting as many as 10,000 military GPS devices.}\\&\multicolumn{5}{p{\dimexpr 5\tabucolX+5\tabcolsep+\arrayrulewidth\relax}}{\textit{Primary/Contemporary References}:~\cite{associatedpressGlitchShowsHow2010}~\textit{Secondary References}:~\cite{zattiProtectionSpaceMissions2017}}\\\\
	
	2010 & Jamming & Government & Iran & Commercial & France\\*
	& \multicolumn{5}{p{\dimexpr 5\tabucolX+5\tabcolsep+\arrayrulewidth\relax}}{A series of jamming incidents around the 31st anniversary of the Islamic Revolution in Iran jammed broadcasts from international satellite television channels on a Eutelsat satellite. The Iranian government is suspected of instigating the attacks.}\\&\multicolumn{5}{p{\dimexpr 5\tabucolX+5\tabcolsep+\arrayrulewidth\relax}}{\textit{Primary/Contemporary References}:~\cite{bbcEUPressuresIran2010}~\textit{Secondary References}:~\cite{fritzSatelliteHackingGuide2013}}\\\\
	
	2010 & Jamming & Government & Jordan & Commercial & United Arab Emirates\\*
	& \multicolumn{5}{p{\dimexpr 5\tabucolX+5\tabcolsep+\arrayrulewidth\relax}}{In 2010 Jordan was accused of jamming Al-Jazeera satellite television feeds including some which broadcast the World Cup.}\\&\multicolumn{5}{p{\dimexpr 5\tabucolX+5\tabcolsep+\arrayrulewidth\relax}}{\textit{Primary/Contemporary References}:~\cite{bbcmonitoringmiddleeastJammingAlJazeeraTV2010}}\\\\
	
	2010 & Jamming & Government & North Korea & Navigational & South Korea\\*
	& \multicolumn{5}{p{\dimexpr 5\tabucolX+5\tabcolsep+\arrayrulewidth\relax}}{North Korea has attempted to disrupt South Korean GPS navigational signals through jamming attacks starting in 2010 and continuing thereafter.}\\&\multicolumn{5}{p{\dimexpr 5\tabucolX+5\tabcolsep+\arrayrulewidth\relax}}{\textit{Primary/Contemporary References}:~\cite{bbcmonitoringasiapacificSouthKoreanSatellite2012}}\\\\
	
	2010 & Groundstation & Government & United States & Gov. Scientific & United States\\*
	& \multicolumn{5}{p{\dimexpr 5\tabucolX+5\tabcolsep+\arrayrulewidth\relax}}{An Office of the Inspector General for NASA audit found that e-waste systems prepared for resale relating to the Space Shuttle missions retained sensitive data which was not correctly deleted, including export controlled information. Similar sensitive information was found on hard drives in dumpster outside a NASA facility.}\\&\multicolumn{5}{p{\dimexpr 5\tabucolX+5\tabcolsep+\arrayrulewidth\relax}}{\textit{Primary/Contemporary References}:~\cite{martinNASACybersecurityExamination2012}}\\\\
	
	2011 & Groundstation & Government & United States & Gov. Scientific & United States\\*
	& \multicolumn{5}{p{\dimexpr 5\tabucolX+5\tabcolsep+\arrayrulewidth\relax}}{The Office of the Inspector General for NASA issued a report indicating that critical vulnerabilities were found in at least six systems which could be used by a remote attacker to control or debilitate ongoing satellite missions.}\\&\multicolumn{5}{p{\dimexpr 5\tabucolX+5\tabcolsep+\arrayrulewidth\relax}}{\textit{Primary/Contemporary References}:~\cite{martinNASACybersecurityExamination2012}}\\\\
	
	2011 & Groundstation & Unknown & China & Gov. Scientific & United States\\*
	& \multicolumn{5}{p{\dimexpr 5\tabucolX+5\tabcolsep+\arrayrulewidth\relax}}{Attackers in 2011 gained administrative control of computer systems in the NASA  Jet Propulsion Laboratory using previously stolen credentials. The attack was later attributed to China.}\\&\multicolumn{5}{p{\dimexpr 5\tabucolX+5\tabcolsep+\arrayrulewidth\relax}}{\textit{Primary/Contemporary References}:~\cite{bbcHackersControlledNasa2012}~\textit{Secondary References}:~\cite{falcoVacuumSpaceCyber2018}}\\\\
	
	2011 & Jamming & Government & Bahrain & Commercial & France\\*
	& \multicolumn{5}{p{\dimexpr 5\tabucolX+5\tabcolsep+\arrayrulewidth\relax}}{A Bahraini opposition station called LuaLua TV was jammed within 5 hours of its first broadcast over a Eutelsat transponder, likely by the Bahraini government.}\\&\multicolumn{5}{p{\dimexpr 5\tabucolX+5\tabcolsep+\arrayrulewidth\relax}}{\textit{Primary/Contemporary References}:~\cite{messiehBahrainSatelliteChannel2011}~\textit{Secondary References}:~\cite{fritzSatelliteHackingGuide2013}}\\\\
	
	2011 & Jamming & Government & Ethiopia & Commercial & United States\\*
	& \multicolumn{5}{p{\dimexpr 5\tabucolX+5\tabcolsep+\arrayrulewidth\relax}}{Ethiopian Satellite Television - an anti-regime satellite television channel - was jammed by the Ethiopian government in 2010 (and several times thereafter). Some have suggested that the equipment and technology for these attacks was provided by Chinese government officials.}\\&\multicolumn{5}{p{\dimexpr 5\tabucolX+5\tabcolsep+\arrayrulewidth\relax}}{\textit{Primary/Contemporary References}:~\cite{ecadfChinaAccusedJamming2011}~\textit{Secondary References}:~\cite{fritzSatelliteHackingGuide2013}}\\\\
	
	2011 & Jamming & Government & Libya & Commercial & United Arab Emirates\\*
	& \multicolumn{5}{p{\dimexpr 5\tabucolX+5\tabcolsep+\arrayrulewidth\relax}}{In 2011 the Libyan government again jammed satellite telephone frequencies in order to combat the use of satphones by smugglers.}\\&\multicolumn{5}{p{\dimexpr 5\tabucolX+5\tabcolsep+\arrayrulewidth\relax}}{\textit{Primary/Contemporary References}:~\cite{thurayapressofficeThurayaTelecomServices2011}~\textit{Secondary References}:~\cite{fritzSatelliteHackingGuide2013}}\\\\
	
	2011 & Jamming & Government & Saudi Arabia & Commercial & Iran\\*
	& \multicolumn{5}{p{\dimexpr 5\tabucolX+5\tabcolsep+\arrayrulewidth\relax}}{Iran has accused Saudi Arabia of jamming its state run satellite television networks starting in 2011.}\\&\multicolumn{5}{p{\dimexpr 5\tabucolX+5\tabcolsep+\arrayrulewidth\relax}}{\textit{Primary/Contemporary References}:~\cite{bbcmonitoringworldmediaIranArabicTV2011}}\\\\
	
	2011 & Groundstation & Unknown & Unknown & Gov. Scientific & United States\\*
	& \multicolumn{5}{p{\dimexpr 5\tabucolX+5\tabcolsep+\arrayrulewidth\relax}}{In 2011 a laptop containing command and control algorithms used for the operation of the International Space Station was stolen. The laptop was unencrypted, but it is unclear if the attacker specifically targeted NASA information.}\\&\multicolumn{5}{p{\dimexpr 5\tabucolX+5\tabcolsep+\arrayrulewidth\relax}}{\textit{Primary/Contemporary References}:~\cite{martinNASACybersecurityExamination2012}~\textit{Secondary References}:~\cite{manulisCyberSecurityNew2020}}\\\\
	
	2011 & Groundstation & Unknown & China & Gov. Scientific & United States\\*
	& \multicolumn{5}{p{\dimexpr 5\tabucolX+5\tabcolsep+\arrayrulewidth\relax}}{Chinese hackers are suspected of having compromised accounts of privileged users at the Jet Propulsion Laboratory which provided attackers with full access to devices on the network.}\\&\multicolumn{5}{p{\dimexpr 5\tabucolX+5\tabcolsep+\arrayrulewidth\relax}}{\textit{Primary/Contemporary References}:~\cite{martinNASACybersecurityExamination2012}}\\\\
	
	2012 & Groundstation & Individual & Romania & Gov. Scientific & United States\\*
	& \multicolumn{5}{p{\dimexpr 5\tabucolX+5\tabcolsep+\arrayrulewidth\relax}}{In February 2012, NASA's Inspector General pressed charges against a Romanian national for intrusions into Jet Propulsion Laboratory computer systems to steal information regarding a scientific sensor for space missions.}\\&\multicolumn{5}{p{\dimexpr 5\tabucolX+5\tabcolsep+\arrayrulewidth\relax}}{\textit{Primary/Contemporary References}:~\cite{martinNASACybersecurityExamination2012}}\\\\
	
	2012 & Groundstation & Individual & Romania & Gov. Scientific & United States\\*
	& \multicolumn{5}{p{\dimexpr 5\tabucolX+5\tabcolsep+\arrayrulewidth\relax}}{In January 2012, the Romanian government arrested a 20-year-old hacker who had compromised both NASA and Romanian government information systems. Other than a low-impact denial of service, this had no lasting repercussions.}\\&\multicolumn{5}{p{\dimexpr 5\tabucolX+5\tabcolsep+\arrayrulewidth\relax}}{\textit{Primary/Contemporary References}:~\cite{martinNASACybersecurityExamination2012}}\\\\
	
	2012 & Jamming & Government & Ertirea & Commercial & France\\*
	& \multicolumn{5}{p{\dimexpr 5\tabucolX+5\tabcolsep+\arrayrulewidth\relax}}{An Eritrean opposition satellite radio channel called Radio Erena was jammed by the Eritrean government in 2012.}\\&\multicolumn{5}{p{\dimexpr 5\tabucolX+5\tabcolsep+\arrayrulewidth\relax}}{\textit{Primary/Contemporary References}:~\cite{leoHerosOrdinaires2013}}\\\\
	
	2012 & Jamming & Government & Ethiopia & Commercial & United Arab Emirates\\*
	& \multicolumn{5}{p{\dimexpr 5\tabucolX+5\tabcolsep+\arrayrulewidth\relax}}{The Ethiopian government is suspected of jamming Eritrean satellite communications signals on ARABSAT platforms starting in 2012 (and several times thereafter).}\\&\multicolumn{5}{p{\dimexpr 5\tabucolX+5\tabcolsep+\arrayrulewidth\relax}}{\textit{Primary/Contemporary References}:~\cite{richardsonEritreaAccusesEthiopia2012}~\textit{Secondary References}:~\cite{fritzSatelliteHackingGuide2013}}\\\\
	
	2012 & Jamming & Government & Syria & Commercial & France\\*
	& \multicolumn{5}{p{\dimexpr 5\tabucolX+5\tabcolsep+\arrayrulewidth\relax}}{Eutelsat was targeted by jamming signals believed to originate in Syria.}\\&\multicolumn{5}{p{\dimexpr 5\tabucolX+5\tabcolsep+\arrayrulewidth\relax}}{\textit{Primary/Contemporary References}:~\cite{bbcmonitoringworldmediaWorldBroadcastersCondemn2012}}\\\\
	
	2012 & Jamming & Government & North Korea & Gov. Military & South Korea\\*
	& \multicolumn{5}{p{\dimexpr 5\tabucolX+5\tabcolsep+\arrayrulewidth\relax}}{North Korea is believed to have jammed South Korean military communications satellites starting in 2012.}\\&\multicolumn{5}{p{\dimexpr 5\tabucolX+5\tabcolsep+\arrayrulewidth\relax}}{\textit{Primary/Contemporary References}:~\cite{bbcmonitoringasiapacificNorthKoreaIncreases2013}}\\\\
	
	2012 & Cryptographic & Researcher & Germany & Multiple & Multiple\\*
	& \multicolumn{5}{p{\dimexpr 5\tabucolX+5\tabcolsep+\arrayrulewidth\relax}}{German researchers published a paper detailing the ability to decrypt voice communications over many satellite phones implementing the common GMR-1 and GMR-2 encryption algorithms.}\\&\multicolumn{5}{p{\dimexpr 5\tabucolX+5\tabcolsep+\arrayrulewidth\relax}}{\textit{Primary/Contemporary References}:~\cite{driessenDonTrustSatellite2012}}\\\\
	
	2013 & Spoofing & Researcher & United States & Navigational & United States\\*
	& \multicolumn{5}{p{\dimexpr 5\tabucolX+5\tabcolsep+\arrayrulewidth\relax}}{In 2009 University of Texas at Austin researchers demonstrated the ability to leverage GPS spoofing to redirect an {\$}80 million yacht remotely.}\\&\multicolumn{5}{p{\dimexpr 5\tabucolX+5\tabcolsep+\arrayrulewidth\relax}}{\textit{Primary/Contemporary References}:~\cite{utaustinpressofficeUTAustinResearchers2013}~\textit{Secondary References}:~\cite{zattiProtectionSpaceMissions2017}}\\\\
	
	2013 & Jamming & Government & Azerbaijan & Commercial & Turkey\\*
	& \multicolumn{5}{p{\dimexpr 5\tabucolX+5\tabcolsep+\arrayrulewidth\relax}}{The Azerbaijani government was found by the USA to be deliberately jamming opposition satellite television stations on Turksat platforms.}\\&\multicolumn{5}{p{\dimexpr 5\tabucolX+5\tabcolsep+\arrayrulewidth\relax}}{\textit{Primary/Contemporary References}:~\cite{bbcmonitoringtranscaucasusunitAzeriEditorSays2013}}\\\\
	
	2013 & Jamming & Government & Egypt & Commercial & Qatar\\*
	& \multicolumn{5}{p{\dimexpr 5\tabucolX+5\tabcolsep+\arrayrulewidth\relax}}{The Egyptian government was accused of jamming Al Jazeera satellite broadcasts during instability in 2013.}\\&\multicolumn{5}{p{\dimexpr 5\tabucolX+5\tabcolsep+\arrayrulewidth\relax}}{\textit{Primary/Contemporary References}:~\cite{trendEgyptJammingJazeera2013}}\\\\
	
	2013 & Jamming & Individual & United States & Navigational & United States\\*
	& \multicolumn{5}{p{\dimexpr 5\tabucolX+5\tabcolsep+\arrayrulewidth\relax}}{A limousine driver in New Jersey had installed a GPS jammer in his vehicle to prevent his employer from tracking the vehicle. The jammer caused interference with navigational systems at a nearby airport.}\\&\multicolumn{5}{p{\dimexpr 5\tabucolX+5\tabcolsep+\arrayrulewidth\relax}}{\textit{Primary/Contemporary References}:~\cite{cbsnewyorkManJamIllegal2013}~\textit{Secondary References}:~\cite{manulisCyberSecurityNew2020}}\\\\
	
	2014 & Groundstation & Official Audit & United States & Gov. Scientific & United States\\*
	& \multicolumn{5}{p{\dimexpr 5\tabucolX+5\tabcolsep+\arrayrulewidth\relax}}{US department of commerce office of the inspector general found more than 9,000 high risk issues in the Joint Polar Satellite System (NOAA) ground stations}\\&\multicolumn{5}{p{\dimexpr 5\tabucolX+5\tabcolsep+\arrayrulewidth\relax}}{\textit{Primary/Contemporary References}:~\cite{crawleyExpeditedEffortsNeeded2014}}\\\\
	
	2014 & Groundstation & Researcher & United States & Commercial & United States\\*
	& \multicolumn{5}{p{\dimexpr 5\tabucolX+5\tabcolsep+\arrayrulewidth\relax}}{A presentation at Defcon in 2014 found severe vulnerabilities  - such as hard-coded passcodes) in 10 SATCOM terminals. Some of these are remotely exploitable but many require physical or at least logical access to the devices.}\\&\multicolumn{5}{p{\dimexpr 5\tabucolX+5\tabcolsep+\arrayrulewidth\relax}}{\textit{Primary/Contemporary References}:~\cite{santamartaSATCOMTerminalsHacking2014}}\\\\
	
	2014 & Jamming & Dissident & Thailand & Commercial & Thailand\\*
	& \multicolumn{5}{p{\dimexpr 5\tabucolX+5\tabcolsep+\arrayrulewidth\relax}}{Thailand government television stations were repeatedly jammed in 2014 during a series of government protestors. No culprit was identified but it is believed to have been the protestors.}\\&\multicolumn{5}{p{\dimexpr 5\tabucolX+5\tabcolsep+\arrayrulewidth\relax}}{\textit{Primary/Contemporary References}:~\cite{istvThailandSuffersSatellite2014}}\\\\
	
	2014 & Jamming & Unknown & Egypt & Commercial & Saudi Arabia\\*
	& \multicolumn{5}{p{\dimexpr 5\tabucolX+5\tabcolsep+\arrayrulewidth\relax}}{In 2014 a comedy broadcast in Egypt was deliberately jammed with interference from two stations in Cairo. It is unclear who is responsible.}\\&\multicolumn{5}{p{\dimexpr 5\tabucolX+5\tabcolsep+\arrayrulewidth\relax}}{\textit{Primary/Contemporary References}:~\cite{apNetworkSignalJammed2014}}\\\\
	
	2014 & Jamming & Government & Libya & Commercial & United Arab Emirates\\*
	& \multicolumn{5}{p{\dimexpr 5\tabucolX+5\tabcolsep+\arrayrulewidth\relax}}{Libya is believed to have jammed a dozen channels by Dubai-headquartered MBC.}\\&\multicolumn{5}{p{\dimexpr 5\tabucolX+5\tabcolsep+\arrayrulewidth\relax}}{\textit{Primary/Contemporary References}:~\cite{ictmonitorMBCHitLibya2014}}\\\\
	
	2014 & Groundstation & Government & China & Gov. Military & Germany\\*
	& \multicolumn{5}{p{\dimexpr 5\tabucolX+5\tabcolsep+\arrayrulewidth\relax}}{Hackers are accused of having compromised computer systems at Deutsche Zentrum f{\~A}{\textonequarter}r Luft- und Raumfahrt (DLR) with spyware that may have been able to implicate the security of critical space missions and missile technologies. Initial attribution suggests Chinese attackers, but the evidence is uncertain.}\\&\multicolumn{5}{p{\dimexpr 5\tabucolX+5\tabcolsep+\arrayrulewidth\relax}}{\textit{Primary/Contemporary References}:~\cite{derspielgelDLRMitTrojanern2014}~\textit{Secondary References}:~\cite{manulisCyberSecurityNew2020}}\\\\
	
	2014 & Eavesdropping & Government & Russia & Multiple & Multiple\\*
	& \multicolumn{5}{p{\dimexpr 5\tabucolX+5\tabcolsep+\arrayrulewidth\relax}}{The Russian satellite Lurch, launched in 2014, is suspected of hovering close to other communications satellites in order to intercept signals}\\&\multicolumn{5}{p{\dimexpr 5\tabucolX+5\tabcolsep+\arrayrulewidth\relax}}{\textit{Primary/Contemporary References}:~\cite{insidesatellitetvRussiaEavesdroppingSatellite2015}}\\\\
	
	2014 & Groundstation & Government & China & Gov. Military & United States\\*
	& \multicolumn{5}{p{\dimexpr 5\tabucolX+5\tabcolsep+\arrayrulewidth\relax}}{In 2014 Crowdstrike released a report indicating that Chinese government-affiliated hackers targeted information about satellite control systems and successfully compromised some sensitive space networks. Few additional details are available.}\\&\multicolumn{5}{p{\dimexpr 5\tabucolX+5\tabcolsep+\arrayrulewidth\relax}}{\textit{Primary/Contemporary References}:~\cite{newmanReportAnotherChinese2014}}\\\\
	
	2014 & Signal Hijacking & Dissident & Palestine & Commercial & Israel\\*
	& \multicolumn{5}{p{\dimexpr 5\tabucolX+5\tabcolsep+\arrayrulewidth\relax}}{Hamas briefly successfully compromised Israeli Channel 10 satellite television broadcasts and transmitted a message threatening Gaza residents.}\\&\multicolumn{5}{p{\dimexpr 5\tabucolX+5\tabcolsep+\arrayrulewidth\relax}}{\textit{Primary/Contemporary References}:~\cite{bbcmonitoringworldmediaHamasHacksSatellite2014}}\\\\
	
	2014 & Groundstation & Government & China & Government - Weather & United States\\*
	& \multicolumn{5}{p{\dimexpr 5\tabucolX+5\tabcolsep+\arrayrulewidth\relax}}{Chinese hackers, believed to be associated with the Chinese government, compromised a sensitive network related to NOAA weather satellites and caused a brief network outage during incident response. It is unclear what systems were compromised or what ability the hackers had.}\\&\multicolumn{5}{p{\dimexpr 5\tabucolX+5\tabcolsep+\arrayrulewidth\relax}}{\textit{Primary/Contemporary References}:~\cite{flahertyChineseHackWeather2014}}\\\\
	
	2014 & Jamming & Government & Ethiopia & Commercial & Saudi Arabia\\*
	& \multicolumn{5}{p{\dimexpr 5\tabucolX+5\tabcolsep+\arrayrulewidth\relax}}{Television broadcasts from the ARABSAT platform were jammed by an attacker in Ethiopia, potenitally associated with the Ethiopean state which has a history of similar jamming attacks targeting Eritrean broadcasts. However, some sources have conjectured that the incident was accidental as ARABSAT does not broadcast to either country.}\\&\multicolumn{5}{p{\dimexpr 5\tabucolX+5\tabcolsep+\arrayrulewidth\relax}}{\textit{Primary/Contemporary References}:~\cite{arabsatArabsatSubjectJamming2014}~\textit{Secondary References}:~\cite{manulisCyberSecurityNew2020}}\\\\
	
	2015 & Signal Injection & Criminal & Russia & Commercial & Multiple\\*
	& \multicolumn{5}{p{\dimexpr 5\tabucolX+5\tabcolsep+\arrayrulewidth\relax}}{Russian-government affiliated group Turla was found to use satellite internet signals to exfiltrate data from malware infections with minimum traceability. Evidence of this method was found in malware dating back to 2007.}\\&\multicolumn{5}{p{\dimexpr 5\tabucolX+5\tabcolsep+\arrayrulewidth\relax}}{\textit{Primary/Contemporary References}:~\cite{tanaseSatelliteTurlaAPT2015}}\\\\
	
	2015 & Eavesdropping & Researcher & United States & Commercial & United States\\*
	& \multicolumn{5}{p{\dimexpr 5\tabucolX+5\tabcolsep+\arrayrulewidth\relax}}{ 2015 Blackhat demonstration indicated practical ability to spoof devices on the Globalstar network and intercept simplex data messages intended for other devices. Globalstar contended that they simply provide hardware and that encryption was the job of their clients based on mission need.}\\&\multicolumn{5}{p{\dimexpr 5\tabucolX+5\tabcolsep+\arrayrulewidth\relax}}{\textit{Primary/Contemporary References}:~\cite{mooreSpreadSpectrumSatcom2015}}\\\\
	
	2015 & Eavesdropping & Researcher & Germany & Commercial & United States\\*
	& \multicolumn{5}{p{\dimexpr 5\tabucolX+5\tabcolsep+\arrayrulewidth\relax}}{Security researchers demonstrated the ability to intercept and interpret communications over the Iridium LEO network using a software defined radio.}\\&\multicolumn{5}{p{\dimexpr 5\tabucolX+5\tabcolsep+\arrayrulewidth\relax}}{\textit{Primary/Contemporary References}:~\cite{secandschneiderIridiumHacking2015}~\textit{Secondary References}:~\cite{manulisCyberSecurityNew2020}}\\\\
	
	2015 & Groundstation & Individual & United Kingdom & Gov. Military & United States\\*
	& \multicolumn{5}{p{\dimexpr 5\tabucolX+5\tabcolsep+\arrayrulewidth\relax}}{A British individual was arrested on charges related to compromising pentagon satellite communications systems. The hacker posted threats online claiming to have the ability to ''control'' satellites but the Pentagon has not confirmed the extent of the intrusion.}\\&\multicolumn{5}{p{\dimexpr 5\tabucolX+5\tabcolsep+\arrayrulewidth\relax}}{\textit{Primary/Contemporary References}:~\cite{deanBritonHeldCyberattacker2015}}\\\\
	
	2016 & Jamming & Government & North Korea & Navigational & South Korea\\*
	& \multicolumn{5}{p{\dimexpr 5\tabucolX+5\tabcolsep+\arrayrulewidth\relax}}{In April 2016 North Korea resumed the 2012 (and occasionally thereafter) jamming campaign against South Korean GPS signals. Russia is suspected (but not proven) to have provided the jamming equipment.}\\&\multicolumn{5}{p{\dimexpr 5\tabucolX+5\tabcolsep+\arrayrulewidth\relax}}{\textit{Primary/Contemporary References}:~\cite{bbcKoreaJammingGPS2016}}\\\\
	
	2016 & Jamming & Government & Russia & Commercial & Ukraine\\*
	& \multicolumn{5}{p{\dimexpr 5\tabucolX+5\tabcolsep+\arrayrulewidth\relax}}{Media Group Ukraine's broadcast of a 2016 football match was targeted by a malicious jamming attack. No attribution for the attack has been made but Russia is highly suspected.}\\&\multicolumn{5}{p{\dimexpr 5\tabucolX+5\tabcolsep+\arrayrulewidth\relax}}{\textit{Primary/Contemporary References}:~\cite{bbcmonitoringkievunitUkrainianTVChannels2016}}\\\\
	
	2016 & Signal Hijacking & Dissident & Palestine & Commercial & Israel\\*
	& \multicolumn{5}{p{\dimexpr 5\tabucolX+5\tabcolsep+\arrayrulewidth\relax}}{Hamas again compromised satellite transmissions, this time of the popular Israeli TV show Big Brother, and replaced them with propaganda films.}\\&\multicolumn{5}{p{\dimexpr 5\tabucolX+5\tabcolsep+\arrayrulewidth\relax}}{\textit{Primary/Contemporary References}:~\cite{housen-courielWhenHamasComes2016}}\\\\
	
	2016 & Signal Hijacking & Individual & Saudi Arabia & Commercial & Israel\\*
	& \multicolumn{5}{p{\dimexpr 5\tabucolX+5\tabcolsep+\arrayrulewidth\relax}}{An individual hacker or group of hackers in Saudi Arabia hijacked Israeli news satellite feeds in protest of an Israeli bill restricting the volume of calls to prayer (muezzin bill). They replaced the media feed with the call to prayer and text threatening punishment from God.}\\&\multicolumn{5}{p{\dimexpr 5\tabucolX+5\tabcolsep+\arrayrulewidth\relax}}{\textit{Primary/Contemporary References}:~\cite{okbiHackersTakeIsraeli2016}}\\\\
	
	2016 & Cryptographic & Researcher & China & Commercial & Multiple\\*
	& \multicolumn{5}{p{\dimexpr 5\tabucolX+5\tabcolsep+\arrayrulewidth\relax}}{Cryptographic researchers in China present a realtime attack against the GMR-2 encryption algorithms used by many satellite phones, updated prior research from Germany.}\\&\multicolumn{5}{p{\dimexpr 5\tabucolX+5\tabcolsep+\arrayrulewidth\relax}}{\textit{Primary/Contemporary References}:~\cite{hudaibSatelliteNetworkHacking2016}}\\\\
	
	2017 & Groundstation & Government & China & Commercial & United States\\*
	& \multicolumn{5}{p{\dimexpr 5\tabucolX+5\tabcolsep+\arrayrulewidth\relax}}{The Chinese Thrip espionage group was found by Symantec in 2017 to have attempted to infect computers which monitor and control satellites.}\\&\multicolumn{5}{p{\dimexpr 5\tabucolX+5\tabcolsep+\arrayrulewidth\relax}}{\textit{Primary/Contemporary References}:~\cite{symantecThripEspionageGroup2017}}\\\\
	
	2017 & Groundstation & Researcher & France & Commercial & United Kingdom\\*
	& \multicolumn{5}{p{\dimexpr 5\tabucolX+5\tabcolsep+\arrayrulewidth\relax}}{A French security researcher on twitter claimed to have compromised a Cobham VSAT terminal on a naval vessel over the internet using a default username and password combination.}\\&\multicolumn{5}{p{\dimexpr 5\tabucolX+5\tabcolsep+\arrayrulewidth\relax}}{\textit{Primary/Contemporary References}:~\cite{chambersShipSatelliteCommunication2017}}\\\\
	
	2018 & Jamming & Government & Israel & Commercial & Syria\\*
	& \multicolumn{5}{p{\dimexpr 5\tabucolX+5\tabcolsep+\arrayrulewidth\relax}}{Israel is suspected of having initiated a jamming attack against Syrian satellite television stations in retaliation for an attack on an Israeli jet flying over Syrian territory.}\\&\multicolumn{5}{p{\dimexpr 5\tabucolX+5\tabcolsep+\arrayrulewidth\relax}}{\textit{Primary/Contemporary References}:~\cite{bbcmonitoringmiddleeastSyrianTVStill2018}}\\\\
	
	2018 & Jamming & Government & Russia & Navigational & NATO\\*
	& \multicolumn{5}{p{\dimexpr 5\tabucolX+5\tabcolsep+\arrayrulewidth\relax}}{Russia is accused of having jammed GPS signals across Norway and Finland to disrupt ongoing NATO war games in the region. The jamming attacks also impacted commercial aviation systems.}\\&\multicolumn{5}{p{\dimexpr 5\tabucolX+5\tabcolsep+\arrayrulewidth\relax}}{\textit{Primary/Contemporary References}:~\cite{thetimesRussiaDisruptedNato2018}}\\\\
	
	2018 & Groundstation & Unknown & Unknown & Gov. Scientific & United States\\*
	& \multicolumn{5}{p{\dimexpr 5\tabucolX+5\tabcolsep+\arrayrulewidth\relax}}{A raspberry-pi microcomputer attached to Jet Propulsion Laboratory systems was compromised and used by attackers to futher access other JPL systems, including systems which control the Deep Space Network radio systems and systems which might allow for malicious control of ongoing space missions.}\\&\multicolumn{5}{p{\dimexpr 5\tabucolX+5\tabcolsep+\arrayrulewidth\relax}}{\textit{Primary/Contemporary References}:~\cite{winderConfirmedNASAHas2019,nasaofficeoftheinspectorgeneralCybersecurityManagementOversight2019}}\\\\
	
	2018 & Groundstation & Unknown & Unknown & Gov. Scientific & United States\\*
	& \multicolumn{5}{p{\dimexpr 5\tabucolX+5\tabcolsep+\arrayrulewidth\relax}}{An advanced persistent threat attacker was found to have compromise Jet Propulsion Laboratory mission networks and to have maintained access to the systems for nearly a year prior to detection in April 2018. They would have had the capability to disable critical space communications systems and were found to have exfiltrated export regulated and sensitive information}\\&\multicolumn{5}{p{\dimexpr 5\tabucolX+5\tabcolsep+\arrayrulewidth\relax}}{\textit{Primary/Contemporary References}:~\cite{nasaofficeoftheinspectorgeneralCybersecurityManagementOversight2019}}\\\\
	
	2018 & Groundstation & Researcher & United States & Multiple & United States\\*
	& \multicolumn{5}{p{\dimexpr 5\tabucolX+5\tabcolsep+\arrayrulewidth\relax}}{An updated version of IOActive research presented in Blackhat 2014 found that VSAT stations could be used to find GPS coordinates of military installations and potentially weaponized to cause interference. The attacks again focused on VSAT terminal firmware.}\\&\multicolumn{5}{p{\dimexpr 5\tabucolX+5\tabcolsep+\arrayrulewidth\relax}}{\textit{Primary/Contemporary References}:~\cite{santamartaLastCallSATCOM2018}}\\\\
	
	2018 & Groundstation & Researcher & Germany & Commercial & Unknown\\*
	& \multicolumn{5}{p{\dimexpr 5\tabucolX+5\tabcolsep+\arrayrulewidth\relax}}{A security researcher demonstrated the ability to compromise maritime satellite terminals over the internet and use them to send NMEA messages to cause harm to operational technology aboard yachts at sea.}\\&\multicolumn{5}{p{\dimexpr 5\tabucolX+5\tabcolsep+\arrayrulewidth\relax}}{\textit{Primary/Contemporary References}:~\cite{gerlingHackingYachtsRemotely2018}}\\\\
	
	2019 & Groundstation & Unknown & Unknown & Gov. Scientific & United States\\*
	& \multicolumn{5}{p{\dimexpr 5\tabucolX+5\tabcolsep+\arrayrulewidth\relax}}{A zero day attack in specialized satellite operations software was compromised on a server belonging to the Jet Propulsion Laboratory. Attackers has the ability to upload control instructions to the spacecraft.}\\&\multicolumn{5}{p{\dimexpr 5\tabucolX+5\tabcolsep+\arrayrulewidth\relax}}{\textit{Primary/Contemporary References}:~\cite{nasaofficeoftheinspectorgeneralCybersecurityManagementOversight2019}}\\\\
	
	2019 & Groundstation & Government & United States & Gov. Military & Iran\\*
	& \multicolumn{5}{p{\dimexpr 5\tabucolX+5\tabcolsep+\arrayrulewidth\relax}}{Media sources have asserted that a series of unexpected failures of Iranian rocket launches in 2019 were the result of a US-government sabotage effort involving either supply-chain compromises or cyber-attacks on Iranian launch vehicles. No official confirmation of these conjectures has been made by either state.}\\&\multicolumn{5}{p{\dimexpr 5\tabucolX+5\tabcolsep+\arrayrulewidth\relax}}{\textit{Primary/Contemporary References}:~\cite{sangerRevivesSecretProgram2019,doffmanCrashedUAEMilitary2019}}\\\\
	
	2019 & Groundstation & Government & Iran & Gov. Military & United Arab Emirates\\*
	& \multicolumn{5}{p{\dimexpr 5\tabucolX+5\tabcolsep+\arrayrulewidth\relax}}{Media sources have asserted that an unexpected launch failure of an military spy satellite belonging to the UAE may have arisen from an Iranian cyber-attack. No official confirmation of these conjectures have been made by either state.}\\&\multicolumn{5}{p{\dimexpr 5\tabucolX+5\tabcolsep+\arrayrulewidth\relax}}{\textit{Primary/Contemporary References}:~\cite{doffmanCrashedUAEMilitary2019}}\\\\
	
	2019 & Misc. & Individual & United States & Commercial & United States\\*
	& \multicolumn{5}{p{\dimexpr 5\tabucolX+5\tabcolsep+\arrayrulewidth\relax}}{US Astronaut Anne McClain was accused of illicitly accessing the bank account of a former partner whilst living aboard the International Space Station. This was widely reported as the first crime accusation against a person in orbit. However the case has not been resolved and, in 2020, the partner who made the accusations was indicted on charges of making false allegations to law enforcement.}\\&\multicolumn{5}{p{\dimexpr 5\tabucolX+5\tabcolsep+\arrayrulewidth\relax}}{\textit{Primary/Contemporary References}:~\cite{mckieNasaAstronautAccessed2019,fieldstadtWomanWhoAccused2020}}\\\\
	
	2019 & Spoofing & Researcher & United States & Navigational & United States\\*
	& \multicolumn{5}{p{\dimexpr 5\tabucolX+5\tabcolsep+\arrayrulewidth\relax}}{A security researcher at Black Hat USA 2019 demonstrated a series of GPS spoofing attacks against an autonomous vehicle. The researcher was able to cause the vehicle to drive off the road by spoofing measurements of its current location.}\\&\multicolumn{5}{p{\dimexpr 5\tabucolX+5\tabcolsep+\arrayrulewidth\relax}}{\textit{Primary/Contemporary References}:~\cite{murrayLegalGNSSSpoofing2019}}\\\\
	
	2019 & Groundstation & Government & North Korea & Gov. Scientific & India\\*
	& \multicolumn{5}{p{\dimexpr 5\tabucolX+5\tabcolsep+\arrayrulewidth\relax}}{A North Korean attributed malware, dubbed ``Dtrack'' was reported to have been found on computer systems belonging to the Indian Space Research Organisation.  Little information regarding the result of the compromise is publicly available, although some media sources surmise it may relate to the concurrent failure of the Chandrayaan 2 Lunar Lander. No evidence of this claim has been provided.}\\&\multicolumn{5}{p{\dimexpr 5\tabucolX+5\tabcolsep+\arrayrulewidth\relax}}{\textit{Primary/Contemporary References}:~\cite{mazoomdaarNotOnlyKudankulam2019}}\\\\
	
	2020 & Eavesdropping & Researcher & United Kingdom & Commercial & United States\\*
	& \multicolumn{5}{p{\dimexpr 5\tabucolX+5\tabcolsep+\arrayrulewidth\relax}}{A security researcher at Black Hat and DEFCON presented research demonstrating that satellite broadband signals could be intercepted by eavesdroppers using inexpensive home-television equipment. They further demonstrated that this impacted the security and privacy of terrestrial, maritime, and aviation customers. The research was based around some prior academic publications.}\\&\multicolumn{5}{p{\dimexpr 5\tabucolX+5\tabcolsep+\arrayrulewidth\relax}}{\textit{Primary/Contemporary References}:~\cite{pavurSecretsSkyPrivacy2019,pavurTaleSeaSky2020,pavurWhispersStars2020,pavurWhispersStarsPractical2020}}\\\\
	
	2020 & Misc. & Individual & United States & Gov. Military & United States\\*
	& \multicolumn{5}{p{\dimexpr 5\tabucolX+5\tabcolsep+\arrayrulewidth\relax}}{The US Air Force and Defense Digital Service hosted ``Hackasat.'' A series of satellite hacking related events and competitions with the goal of increasing technical exposure to satellite cyber-security. The final challenge of the competition was to upload a mission plan to a live satellite (after exploiting a series of vulnerabilities in a ground-based system meant to replicate a satellite) and take a ``cyber moon-shot'' photograph of the moon using the satellite's onboard camera.}\\&\multicolumn{5}{p{\dimexpr 5\tabucolX+5\tabcolsep+\arrayrulewidth\relax}}{\textit{Primary/Contemporary References}:~\cite{defensedigitalserviceHackasat2020}}\\\\

\end{longtabu}
%\end{landscape}
\todos
\end{document}